\documentclass[prc,floatfix,groupedaddress,nofootinbib,twocolumn,showpacs,preprintnumbers,
amsmath,amssymb,amsfonts,superscriptaddress,widetable] {revtex4}
\usepackage{bm}
\usepackage{mathrsfs}
\usepackage{amssymb}
\usepackage{amsmath}
\usepackage{graphicx}
\usepackage{array}
\usepackage{color}
\newcommand{\rhoz}{\rho_{\raisebox{-0.75pt}{\tiny\!sat}}}
\newcommand{\epsz}{\varepsilon_{\raisebox{-0.75pt}{\tiny 0}}}

\begin{document}
\title{Nuclear Astrophysics in the New Era of Multimessenger Astronomy}
\author{J. Piekarewicz}\email{jpiekarewicz@fsu.edu}
\affiliation{Department of Physics, Florida State University,
               Tallahassee, FL 32306, USA}
\date{\today}
\begin{abstract}
Neutron stars are unique cosmic laboratories for the exploration of
matter under extreme conditions of density and neutron-proton
asymmetry. Due to their enormous dynamic range, neutron stars display
a myriad of exotic states of matter that are impossible to recreate
under normal laboratory conditions. In these three lectures I will
discuss how the strong synergy that has developed between nuclear
physics and astrophysics will uncover some of the deepest secrets
behind these fascinating objects. In particular, I will highlight the
enormous impact that the very first detection of gravitational waves
from the binary neutron-star merger GW170817 is having in
constraining the composition, structure, and dynamics of neutron
stars.
\end{abstract}
\smallskip
\pacs{
04.40.Dg,   
21.65.Ef,   
24.10.Jv,   
26.60.Kp,   
97.60.Jd   
}
\maketitle

\section{Preface}
\label{sec:preface}

Massive stars use the raw materials (mostly hydrogen and helium)
created during the Big Bang to fuel the stars and to create via 
thermonuclear fusion many of the chemical elements found in the
periodic table. However, the fusion of light nuclei into ever
increasing heavier elements terminates abruptly with the synthesis of
the iron-group elements that are characterized by having the largest
binding energy per nucleon. Once the iron core exceeds a
characteristic mass limit of about 1.4 solar masses, neither
thermonuclear fusion nor electron degeneracy pressure can halt the
collapse of the stellar core. The unimpeded collapse of the core and the ensuing
shock wave produce one of the most spectacular events in the Universe:
a Supernova Explosion. Core-collapse supernovae leave behind exotic
compact remnants in the form of either black holes or neutron
stars. Neutron stars are the central theme of the present lectures.

The historical first detection of gravitational waves from the binary
neutron-star (BNS) merger GW170817 by the LIGO-Virgo
collaboration\,\cite{Abbott:PRL2017} is providing fundamental new
insights into the nature of dense matter and the astrophysical site
for the creation of the heavy elements via the rapid neutron-capture
process ($r$-process). Although GW170817 represents the very first
detection of a BNS merger, it is already furnishing answers to two of
the ``eleven science questions for the next century" identified by the
National Academies Committee on the Physics of the
Universe\,\cite{QuarksCosmos:2003}: \emph{What are the new states of
matter at exceedingly high density and temperature?}  and \emph{how
were the elements from iron to uranium made?} In these three lectures
I will try to illuminate the deep connections that exist between
nuclear physics and astrophysics in understanding the composition,
structure, and dynamics of neutron stars. I will discuss how the
combination of nuclear physics insights, modern theoretical
approaches, laboratory experiments, and astronomical observations
using both electromagnetic and gravitational radiation pave the way to
our understanding of these fascinating objects at the dawn of the
brand new era of ``multimessenger" astronomy.

The lectures were divided into three independent units that were aimed 
to provide a coherent picture of the field. In turn, this proceedings are
also divided into three chapters. First, I will provide a historical
perspective that introduces some of the main actors responsible for
the development of the field. Second, I will provide a description of
the many phases and exotic states of matter that we believe ``hide" in
the interior of a neutron star. Finally, I will end by discussing the
deep connections between ``\emph{Heaven and Earth}", namely, the
ongoing and future suite of terrestrial experiments and astronomical
observations that---with appropriate theoretical insights---will unlock
some of the deepest secrets lurking within neutron stars.

\section{Historical Perspective}
\label{sec:historical}

We start this chapter by highlighting the 1939 work by Oppenheimer and
Volkoff, a theoretical milestone in the history of neutron
stars\,\cite{Opp39_PR55}. By then, Einstein's general theory of
relativity was firmly established. On the other hand, the
existence of the neutron was experimentally confirmed just a few years
earlier, in 1932, by James Chadwick working at the Cavendish
Laboratory in the UK\,\cite{Chadwick:1932}. Yet soon after Chadwick's
discovery, the term \emph{neutron star} seems to appears
in writing for the first time in the 1933 proceedings of the the
American Physical Society by Baade and
Zwicky\,\cite{Baade:1934}. Using what it is now commonly referred to
as the Tolman-Volkoff-Oppenheimer (TOV) equations, Oppenheimer and
Volkoff concluded that a neutron star supported exclusively by the
pressure from its degenerate neutrons will collapse once its mass
exceeds 0.7 solar masses (0.7$M_{\odot}$). Unbeknownst to them, this
finding will eventually promote nuclear physics to the forefront of
neutron-star structure---given that neutron stars with masses of at 
least 2$M_{\odot}$ have already been 
observed\,\cite{Demorest:2010bx,Antoniadis:2013pzd}.  In essence, 
the large discrepancy between observation and the theoretical prediction 
by Oppenheimer and Volkoff has transferred ownership of the 
neutron-star problem to nuclear physics.

\begin{figure}[h]
 \vspace{-0.1cm}
 \begin{center}
\includegraphics[width=0.8\columnwidth]{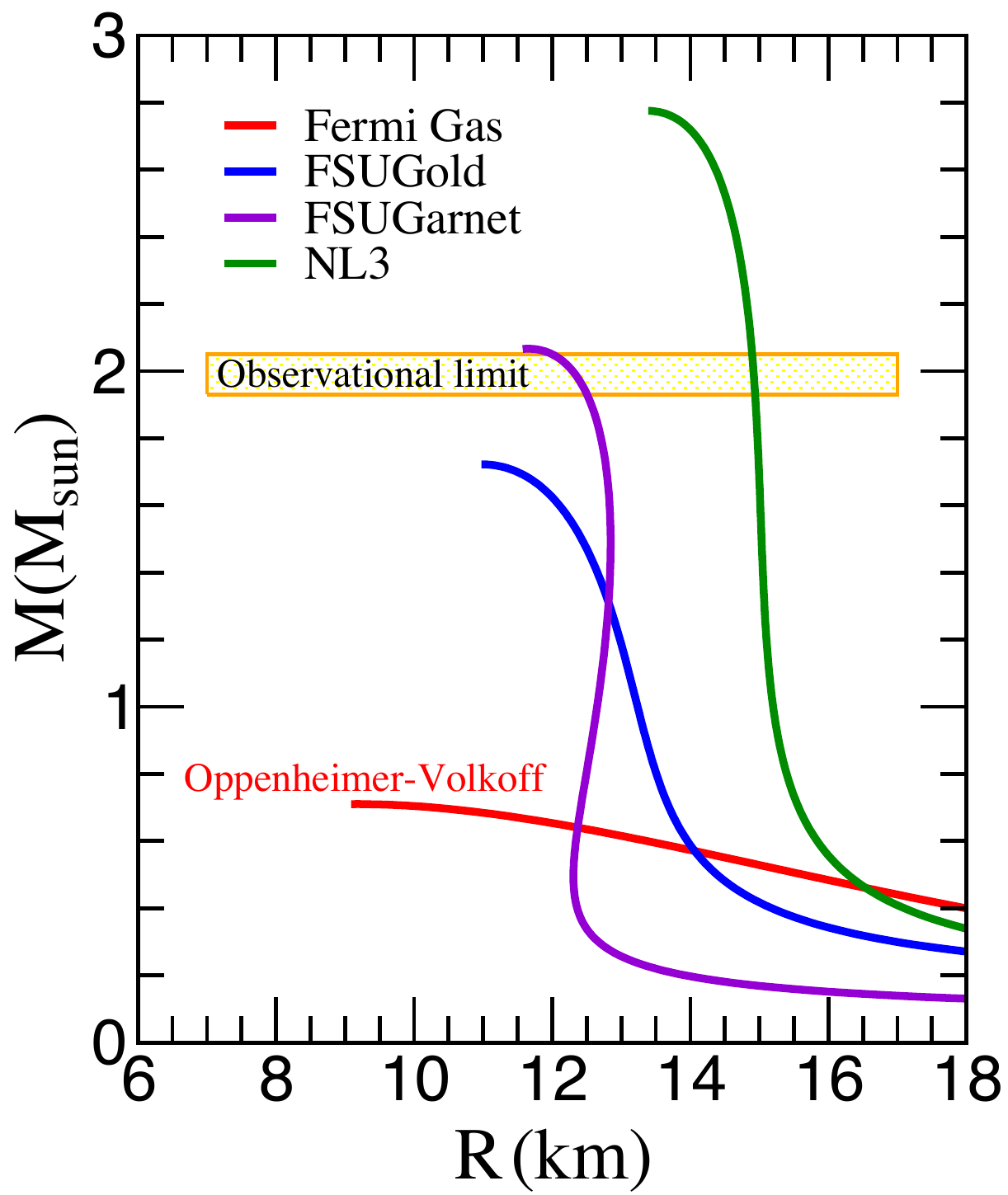}
  \caption{Mass-Radius relation as predicted by a simple
  Fermi-gas model\,\cite{Opp39_PR55} and three realistic 
  equations of state that will be introduced in later chapters. 
  The horizontal band indicates the current observational limit 
  on the maximum stellar mass\,\cite{Demorest:2010bx,
  Antoniadis:2013pzd}.}
 \label{Fig1}
 \end{center} 
 \vspace{-0.25cm}
\end{figure}

The Tolman-Oppenheimer-Volkoff equations, which represent 
a generalization of Newtonian gravity to the realm of general relativity, 
are expressed as a coupled set of first-order differential equations of 
the following form:
 \begin{eqnarray}
  &&\frac{dP}{dr}\!=\!-\!G\,\frac{{\cal E}(r)M(r)}{r^{2}}  
         \left[1\!+\!\frac{P(r)}{{\cal E}(r)}\right]
         \left[1\!+\!\frac{4\pi r^{3}P(r)}{M(r)}\right]
         \nonumber \\ && \hspace{16pt} \times
         \left[1\!-\!\frac{2GM(r)}{r}\right]^{-1}\hspace{-12pt},
         \label{TOVa}  \\
   &&\frac{dM}{dr}=4\pi r^{2}{\cal E}(r)\;,
         \label{TOVb}
 \label{TOV}
\end{eqnarray}
where $G$ is Newton's gravitational constant and $P(r)$, ${\cal
E}(r)$, and $M(r)$ represent the pressure, energy density, and
enclosed-mass profiles of the star, respectively. The three terms
enclosed in square brackets encode the relevant corrections to
Newtonian gravity. The solution to
these equations by Oppenheimer and Volkoff under the assumption that
the equation of state (the relation between the pressure and the energy
density) is that of a free Fermi gas of neutrons yields a maximum
neutron star mass of 0.7$M_{\odot}$.  Note that the fact that the
equation of state (EOS) is the only input that neutron stars are
sensitive to creates a unique synergy between nuclear physics and
astrophysics. The mass-versus-radius relation obtained by Oppenheimer
and Volkoff is displayed with a red solid line in Fig.\ref{Fig1},
alongside the current observational limit on the maximum neutron-star
mass\,\cite{Demorest:2010bx,Antoniadis:2013pzd}. Also shown are
predictions from more realistic models that will be discussed later
and that take into account the complicated and subtle nuclear 
dynamics.

I would be remiss if I did not highlight the indirect, yet pivotal,
role that Subramanyan Chandrasekhar (``Chandra'') played in the
history of neutron stars. Already in 1926 R.H. Fowler---Dirac's
doctoral advisor---showed that white-dwarf stars are supported against
gravitational collapse by quantum degeneracy pressure, the pressure
exerted by a cold Fermi gas by virtue of the Pauli exclusion
principle. In particular, Fowler showed that the electron degeneracy
pressure scales as the $5/3$ power of the electronic density. However,
during his 1930 journey to Cambridge to pursue his doctoral degree
under the supervision of no other than Fowler, Chandrasekhar realized
that as the stellar density increases and the electrons become
relativistic, the pressure support weakens, ultimately becoming
proportional to the $4/3$ power of the electronic density. This
weakening has dramatic consequences: a white-dwarf star with a mass in
excess of about 1.4 solar masses---the so-called ``Chandrasekhar mass
limit''---will collapse under its own
weight\,\cite{Chandrasekhar:1931}. Although such far-reaching result
is now well accepted, at that time it was the subject of
derision, primarily by Arthur Eddington. It is worth noting that
Chadwick's discovery of the neutron came a year after Chandra's
prediction of the Chandrasekhar mass limit. Ultimately, however,
Chandra prevailed and in recognition to his many scientific
contributions NASA launched in 1999 the ``Chandra X-ray Observatory",
NASA's flagship mission for X-ray astronomy. We note that in a 1932
publication Landau---independently of Chandrasekhar---predicts the
existence of a maximum mass for a white dwarf
star\,\cite{Landau:1932}.  Moreover, Landau went ahead to speculate on
the existence of dense stars that look like giant atomic nuclei. For
further historic details see Ref.\,\cite{Yakovlev:2012rd} and chapter
14 in Ref.\,\cite{Meng:2016}.

Although firmly established on theoretical grounds, it would take
almost 30 years after the work by Oppenheimer and Volkoff to
discover neutron stars. The glory of the discovery fell upon the
talented young graduate student Jocelyn Bell, now
\emph{Dame} Jocelyn Bell Burnell. While searching for signals from 
the recently discovered and to this day still enigmatic quasars, Bell
detected a ``bit of scruff'' in the data arriving into her newly
constructed radio telescope. The arriving signal was ``pulsing'' with
such an enormous regularity, 1.337\,302\,088\,331 seconds, that both
Bell and Anthony Hewish (her doctoral advisor) were bewildered by the
detection. Initially convinced that the signal was a beacon from an
extraterrestrial civilization, they dubbed the source as ``Little
Green Man 1''. Now known as radio pulsar ``PSR B1919+21'', Bell had
actually made the very first detection of a rapidly rotating neutron
star\,\cite{Hewish:1968}.  Although it is well known that Jocelyn Bell
was snubbed by the Nobel committee in 1974---the year that her
doctoral advisor Anthony Hewish shared the Nobel prize in Physics with
Martin Ryle---she has always displayed enormous grace and humility in
the face of this controversy. Since then, Bell has been recognized
with an enormous number of honors and awards. Moreover, Dr. Iosif
Shklovsky---a recipient of the 1972 Bruce Medal for outstanding
lifetime contributions to astronomy---paid her one of the highest
compliments that one can receive from a fellow scientist: ``\emph{Miss
Bell, you have made the greatest astronomical discovery of the
twentieth century."}

\section{Anatomy of a Neutron Star}
\label{sec:anatomy}

\begin{figure}[h]
\begin{center}
\includegraphics[width=0.475\columnwidth,height=4cm]{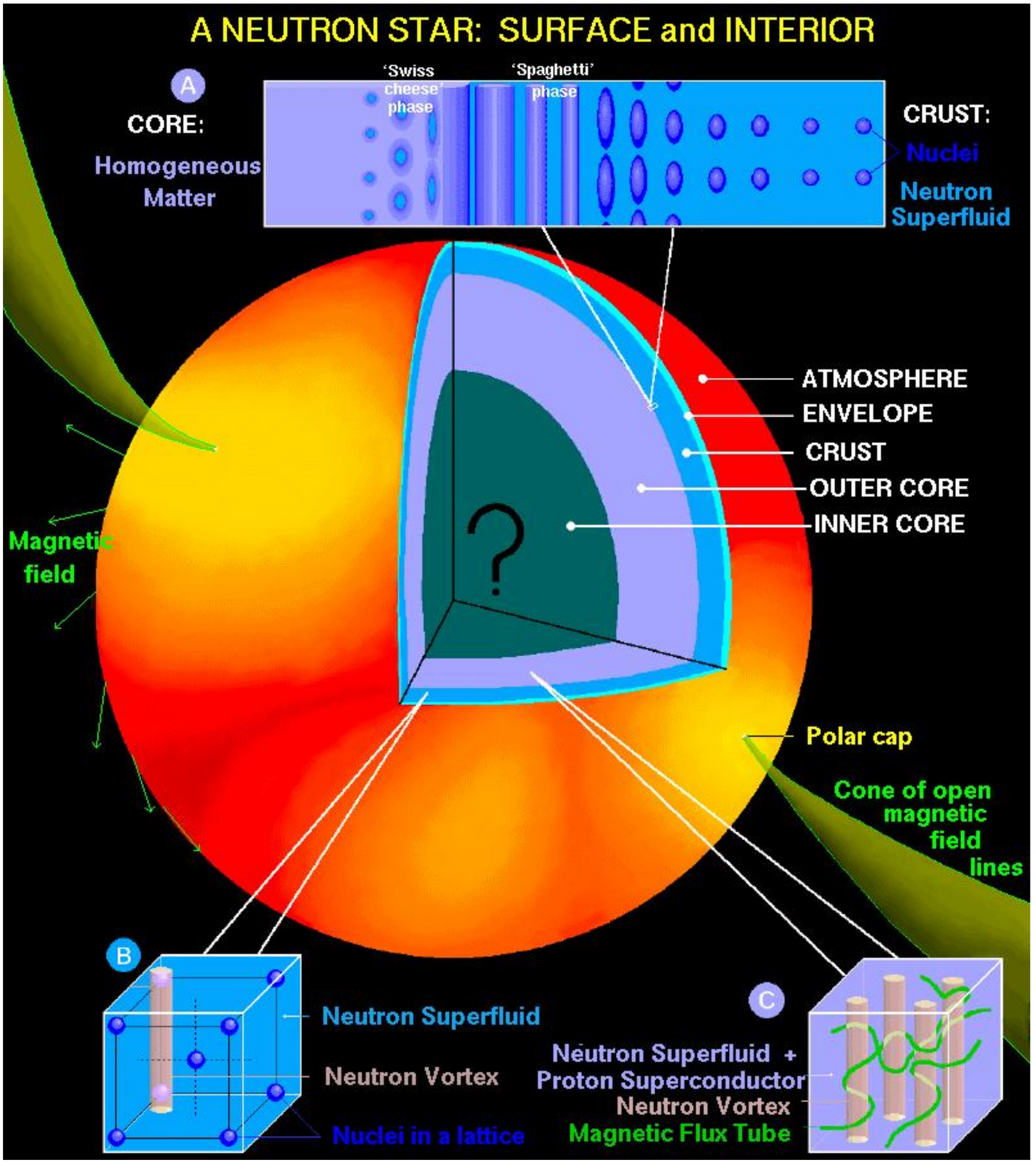}
\includegraphics[width=0.475\columnwidth,height=4cm]{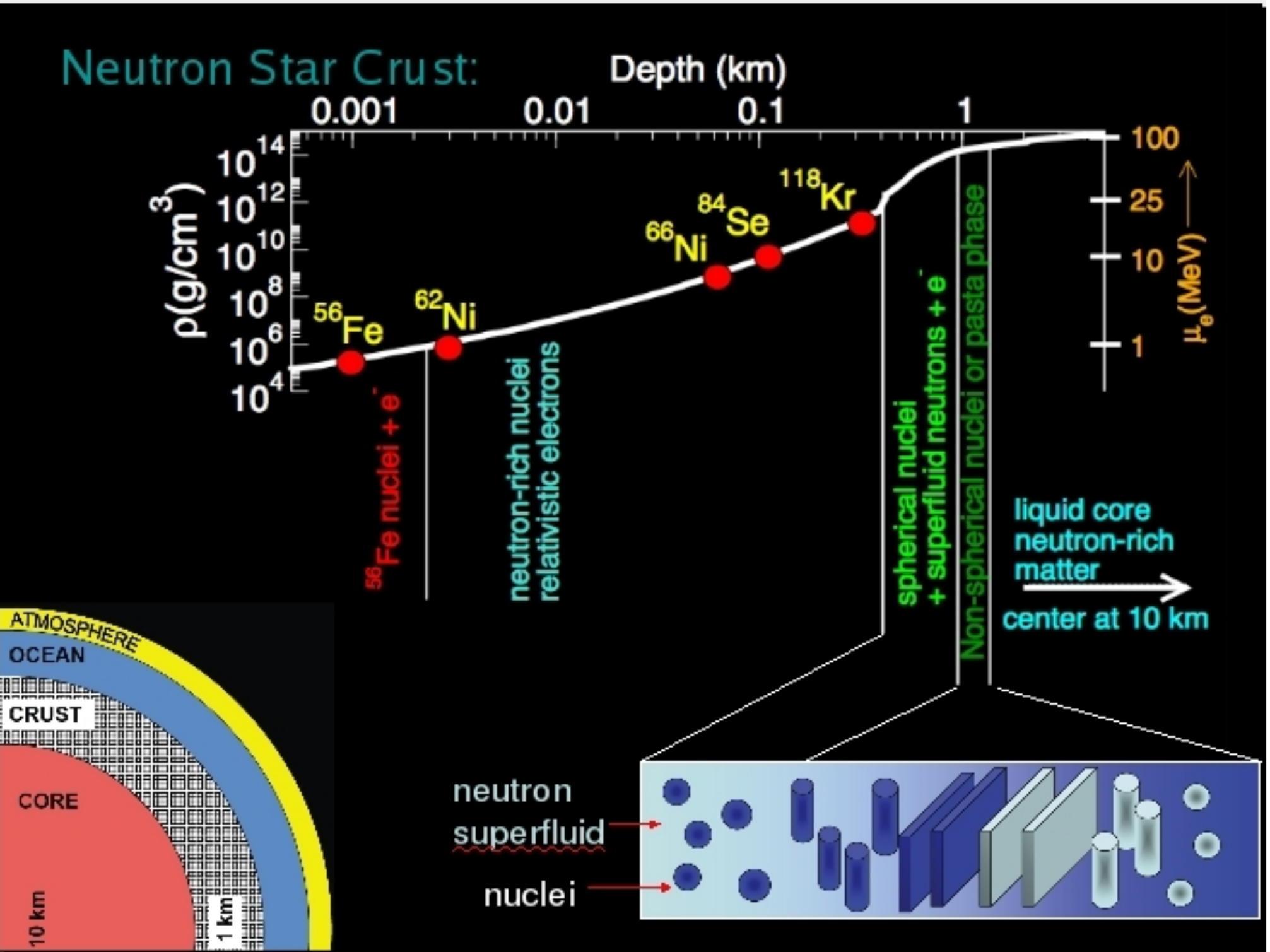}
\caption{The left-hand panel depicts what is believed to be an accurate 
rendition of the fascinating structure and exotic phases that exist in a 
neutron star (courtesy of Dany Page). On the right-hand panel we
display the assumed composition of the crust of a neutron star---from
a crystalline lattice of exotic neutron-rich nuclei to the emergence of
the nuclear-pasta phase (courtesy of Sanjay Reddy).}             
 \label{Fig2}
\end{center}
\end{figure}

The structure of neutron stars is both interesting and complex. To
appreciate the enormous dynamic range and richness displayed by these
fascinating objects, we display in Fig.\,\ref{Fig2} what is believed
to be an accurate rendition of the structure and composition of a
neutron star. Further, to accentuate some of the extreme conditions
present in a neutron star, we display in Table\,\ref{Table1} some of
the characteristic of the Crab pulsar, the compact remnant of a 
supernovae explosion in the constellation Taurus that was observed
nearly 1,000 years ago.

The outermost surface of the neutron star contains a
very thin atmosphere of only a few centimeters thick that is composed
of hydrogen, but may also contain heavier elements such as helium and
carbon. The detected electromagnetic radiation may be used to
constrain critical parameters of the neutron star. For example,
assuming pure blackbody emission from the stellar surface at a
temperature $T$ provides a determination of the stellar radius from
the Stefan-Boltzmann law: $L\!=\!4\pi\sigma
R^{2}T^{4}$. Unfortunately, complications associated with distance
measurements and distortions of the black-body spectrum make the
accurate determination of stellar radii---one of the most critical
observables informing the equation of state---a challenging task. Just
below the atmosphere lies the $\sim\!\!100$\,m thick envelope that acts
as ``blanket'' between the hot interior (with $T\!\gtrsim\!10^{8}$\,K)
and the ``cold'' surface (with
$T\!\gtrsim\!10^{6}$\,K)\,\cite{Page:2004fy}. Further below lies the
non-uniform crust, a region characterized by fascinating exotic states
of matter that are impossible to recreate under normal laboratory
conditions. The non-uniform crust sits above a uniform liquid core
that consists of neutrons, protons, electrons, and muons. The core
accounts for practically all the mass and for about 90\% of the size
of a neutron star. Finally, depending on the highest densities that
may be attained in the inner core, there is also a possibility (marked
with a question mark in Fig.\,\ref{Fig2}) for the emergence of new
exotic phases, such as pion or kaon
condensates\,\cite{Ellis:1995kz,Pons:2000xf}, strange quark
matter\,\cite{Weber:2004kj,Page:2006ud}, and color
superconductors\,\cite{Alford:1998mk,Alford:2007xm}.

\begin{table}[!htbp]
{\begin{tabular}{@{}ll@{}} \toprule
  Name: PSR B0531+21 & Constellation: Taurus\\
  Distance: 2.2 kpc  & Age: 960 years\\
  Mass: $1.4\,M_{\odot}$ & Radius: 10\,km \\ 
  Density: $10^{15}$g/cm${}^{3}$ & Pressure: $10^{29}$\,atm \\  
  Temperature: $10^{6}$\,K & Escape velocity: 0.6\,c\\
  Period: 33\,ms & Magnetic Field: $10^{12}$ G \\ \botrule
\end{tabular}}  
\caption{Characteristics of the 960 year old Crab pulsar.}
\label{Table1}
\end{table}    

\subsection{The Outer Crust: Extreme Sensitivity to Nuclear Masses}
\label{sec:outercrust}

The range of the short-range nucleon-nucleon (NN) interaction is
approximately equal to the Compton wavelength of the pion, or about
$1.4\,{\rm fm}$. In turn, in uniform nuclear matter at \emph{saturation 
density} ($\rhoz\!\approx\!0.15\,{\rm fm}^{3}$) the average 
inter-nucleon separation is about $1.2\,{\rm fm}$. This suggests that 
at densities of about $1/2$ to $1/3$ of nuclear-matter saturation 
density, the average inter-nucleon separation will be large enough 
that the NN interaction will cease to be effective. Thus, to maximize 
the impact of the NN attraction it becomes energetically favorable for 
the system to break translational invariance and for the nucleons to 
cluster into nuclei. The question of which nucleus is energetically the 
most favorable emerges from a dynamical competition between the 
symmetry energy---which favors nearly symmetric nuclei 
($N\!\gtrsim\!Z$)---and the electronic density, which in turn favors 
no electrons ($Z\!=\!0$).

Although subtle, the dynamics of the outer crust is encapsulated in a 
relatively simple expression for the total Gibbs free energy per nucleon, 
which at zero temperature equals the total chemical potential of the 
system. That is\,\cite{Haensel:1989,Haensel:1993zw,Ruester:2005fm,
RocaMaza:2008ja,RocaMaza:2011pk},
 \begin{equation}
   \mu(Z,A; P) = \frac{M(Z,A)}{A} + \frac{Z}{A}\mu_{e} -
   \frac{4}{3}C_{l}\frac{Z^2}{A^{4/3}}\,p_{{}_{\rm F}}. 
   \label{ChemPot}
\end{equation}
The first term is independent of the pressure---or equivalently of the
baryon density---and  represents the entire nuclear 
contribution to the chemical potential. It depends exclusively on 
the mass per nucleon of the ``optimal'' nucleus populating the crystal 
lattice. The second term represents the electronic contribution and, as 
any Fermi gas, it is strongly density dependent.  Finally, the last 
term provides the relatively modest---although by no means 
negligible---electrostatic lattice contribution 
(with $C_{l}\!=\!3.40665\!\times\!10^{-3}$). Here $p_{{}_{\rm F}}$ is 
the nuclear Fermi energy that is related to the baryon density through 
the following expression: 
 \begin{equation}
   p_{{}_{\rm F}} = \left(3\pi^2\rho\right)^{1/3}.
   \label{FermiMom}
\end{equation}
Finally, the connection between the pressure and the baryon 
density is provided by the underlying crustal equation of state 
that, as anticipated, is dominated by the relativistic electrons. 
That is,
 \begin{eqnarray}
   && P(\rho) \!=\! 
   \frac{m_{e}^{4}}{3\pi^{2}} 
   \left(x_{{}_{\rm F}}^{3}y_{{}_{\rm F}} \!-\! 
   \frac{3}{8}\left[x_{{}_{\rm F}}y_{{}_{\rm F}}
   \Big(x_{{}_{\rm F}}^{2}\!+\!y_{{}_{\rm F}}^{2}\Big)
   \!-\! \ln(x_{{}_{\rm F}}\!+\!y_{{}_{\rm F}}) \right]\right)
   \nonumber \\ && \hspace{25pt}
   \!-\! \frac{\rho}{3}C_{l}\frac{Z^2}{A^{4/3}}\,p_{{}_{\rm F}},
 \label{Pressure}   
\end{eqnarray}
where $x_{{}_{\rm F}}\!=\!p_{{}_{\rm F}}^{e}/m_{e}$ and 
$y_{{}_{\rm F}}\!=\!(1\!+\!x_{{}_{\rm F}}^{2})^{1/2}$ are 
scaled electronic Fermi momentum and Fermi energy, 
respectively; $p_{{}_{\rm F}}^{e}\!=\!(Z/A)^{1/3}p_{{}_{\rm F}}$.
This discussion suggests that the only unknown in the determination
of the crustal composition is the optimal nucleus, namely, the one that
minimizes the chemical potential, at a given pressure.

The search for the optimal nucleus is performed as follows. For a given 
pressure $P$ and nuclear species ($Z,A$), the equation of state is used 
to determine the corresponding baryon density of the system which, in turn, 
determines the Fermi momentum $p_{{}_{\rm F}}$ and the electronic 
chemical potential $\mu_{e}$. This is sufficient to compute the chemical 
potential of the system as indicated in Eq.\,(\ref{ChemPot}). However, this 
procedure requires scanning over an entire mass table---which in some instances 
consists of nearly 10,000 nuclei. The $(Z,A)$ combination that minimizes 
$\mu(A,Z; P)$ determines the optimal nucleus populating the crystal lattice 
at the given pressure. Naturally, if the pressure (and thus the density) is 
very small so that the electronic contribution to the chemical potential 
is negligible, then ${}^{56}$Fe---with the lowest mass per nucleon---becomes 
the nucleus of choice. As the pressure and density increase so that the electronic 
contribution may no longer be neglected, then it becomes advantageous 
to reduce the electron fraction $Z/A$ at the expense of increasing the 
neutron-proton asymmetry. This results in an increase in the mass 
per nucleon. Which nucleus becomes the optimal choice then emerges 
from a subtle competition between the electronic contribution that favors 
$Z\!=\!0$ and the nuclear symmetry energy which favors (nearly) symmetric 
nuclei. 

Even though the underlying physics is relatively simple, computing the 
composition of the outer crust is hindered by the unavailability of mass 
measurements of exotic nuclei with a very large neutron-proton asymmetry.
Indeed, of the masses of the $N\!=\!82$ isotones believed to populate the 
deepest layers of the outer crust---such as ${}^{122}$Zr, ${}^{120}$Sr, and 
${}^{118}$Kr---none have been determined 
experimentally\,\cite{AME:2012,AME:2016} and it is unlikely that they 
will ever be determined even at the most powerful rare-isotope facilities. 
Thus, the only recourse is to resort to theoretical calculations which, in
turn, must rely on extrapolations far away from their region of applicability.
Whereas no clear-cut remedy exists to cure such unavoidable 
extrapolations, we have recently offered a path to mitigate the 
problem\,\cite{Utama:2015hva,Utama:2016rad,Utama:2017wqe,
Utama:2017ytc}. The basic paradigm behind our two-pronged approach 
is to start with a robust underlying theoretical model that captures as much 
physics as possible, followed by a \emph{Bayesian Neural Network} (BNN) 
refinement of the \emph{mass residuals} that aims to account for the 
missing physics\,\cite{Utama:2015hva}. That is, the resulting mass 
formula is given by
\begin{equation}
 M(Z,N) \equiv M_{\rm model}(Z,N) + \delta_{\rm model}(Z,N),
 \label{BNN_LDM}
\end{equation}
where $M_{\rm model}(Z,N)$ is the ``bare" model prediction and 
$\delta_{\rm model}(Z,N)$ the BNN refinement to the difference
between the model predictions and experiment. For further details
on the implementation see\,\cite{Utama:2015hva,Utama:2016rad,
Utama:2017wqe,Utama:2017ytc}.

\begin{figure}[ht]
\vspace{-0.05in}
\includegraphics[width=.95\columnwidth,angle=0]{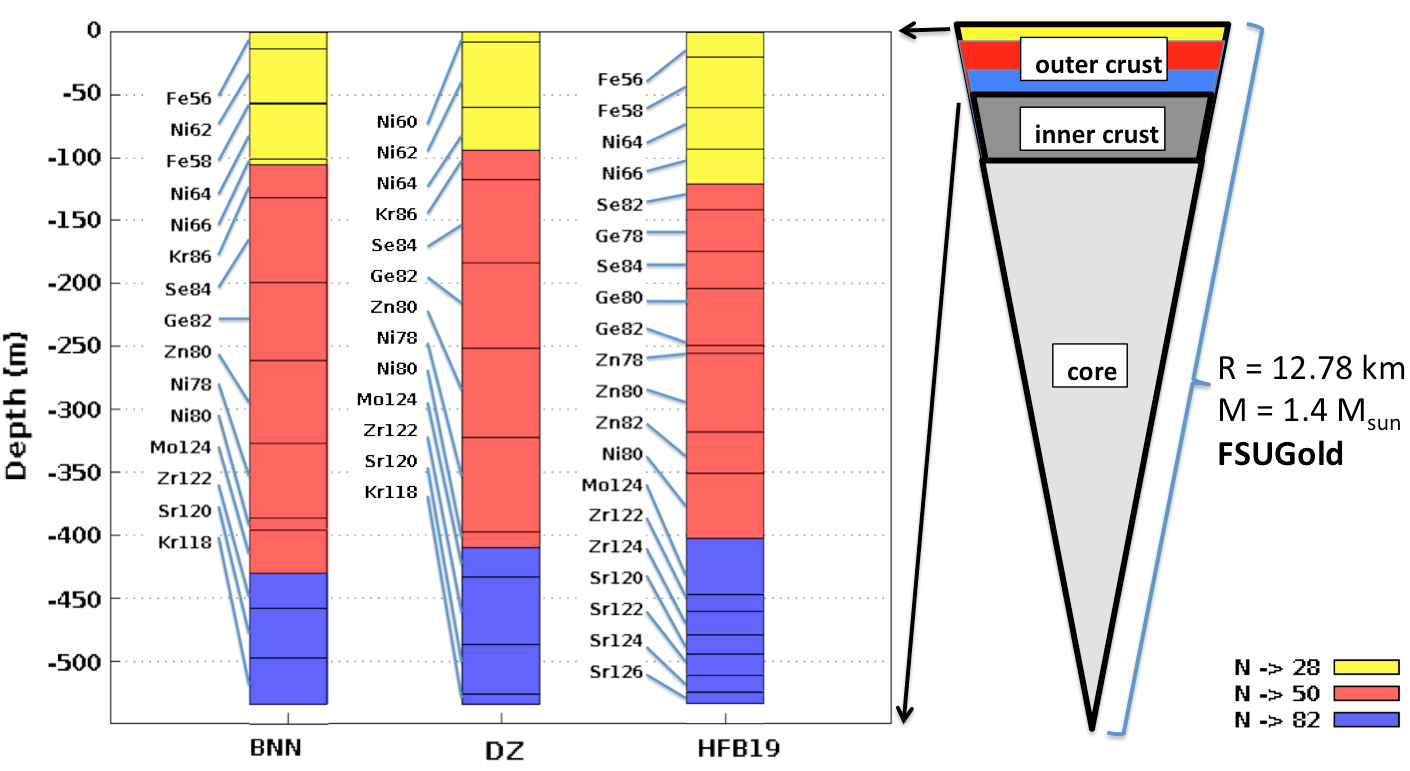}
\caption{Crustal composition of a canonical $1.4\,M_{\odot}$ 
neutron star with a 12.78\,km radius as predicted by three mass 
models: BNN, DZ, and HFB19. See text for further explanations.} 
\label{Fig3}
\end{figure}

In Fig.\,\ref{Fig3} we display the composition of the outer crust as a
function of depth for a neutron star with a mass of $1.4\,M_{\odot}$ 
and a radius of 12.78\,km. Predictions are shown using our newly 
created mass model ``BNN'', Duflo Zuker, and HFB19; these 
last two without any BNN refinement. The composition of the 
upper layers of the crust (spanning about 100 m and depicted in 
yellow) consists of Fe-Ni nuclei with masses that are well known 
experimentally. As the Ni-isotopes become progressively more 
neutron rich, it is energetically favorable to transition into 
the magic $N\!=\!50$ region. In the particular case of the 
BBN-improved model, this intermediate region is 
predicted to start with stable ${}^{86}$Kr and then progressively 
evolve into the more exotic isotones ${}^{84}$Se ($Z\!=\!34$), 
${}^{82}$Ge ($Z\!=\!32$), ${}^{80}$Zn ($Z\!=\!30$), and 
${}^{78}$Ni ($Z\!=\!28$); all this in an effort to reduce the electron 
fraction. In this region, most of the masses are experimentally known, 
although for some of them the quoted value is not derived from purely 
experimental data\,\cite{AME:2012}. Ultimately, it becomes energetically 
favorable for the system to transition into the magic $N\!=\!82$ 
region. In this region \emph{none of the relevant nuclei have 
experimentally determined masses}. Although not shown, it is interesting
to note that the composition of the HFB19 model changes considerably
after the BNN refinement, bringing it into closer agreement with the
predictions of both BNN and Duflo-Zuker. Although beyond the 
scope of this work, we should mention that the crustal composition 
is vital in the study of certain elastic properties of the crust, such as its 
shear modulus and breaking strain---quantities that are of great 
relevance to magnetar starquakes\,\cite{Piro:2005jf,Steiner:2009yg} 
and continuous gravitational-wave emission from rapidly rotating 
neutron stars\,\cite{Horowitz:2009ya}. 

\begin{figure}[h]
\vspace{-0.05in}
\includegraphics[width=0.9\columnwidth]{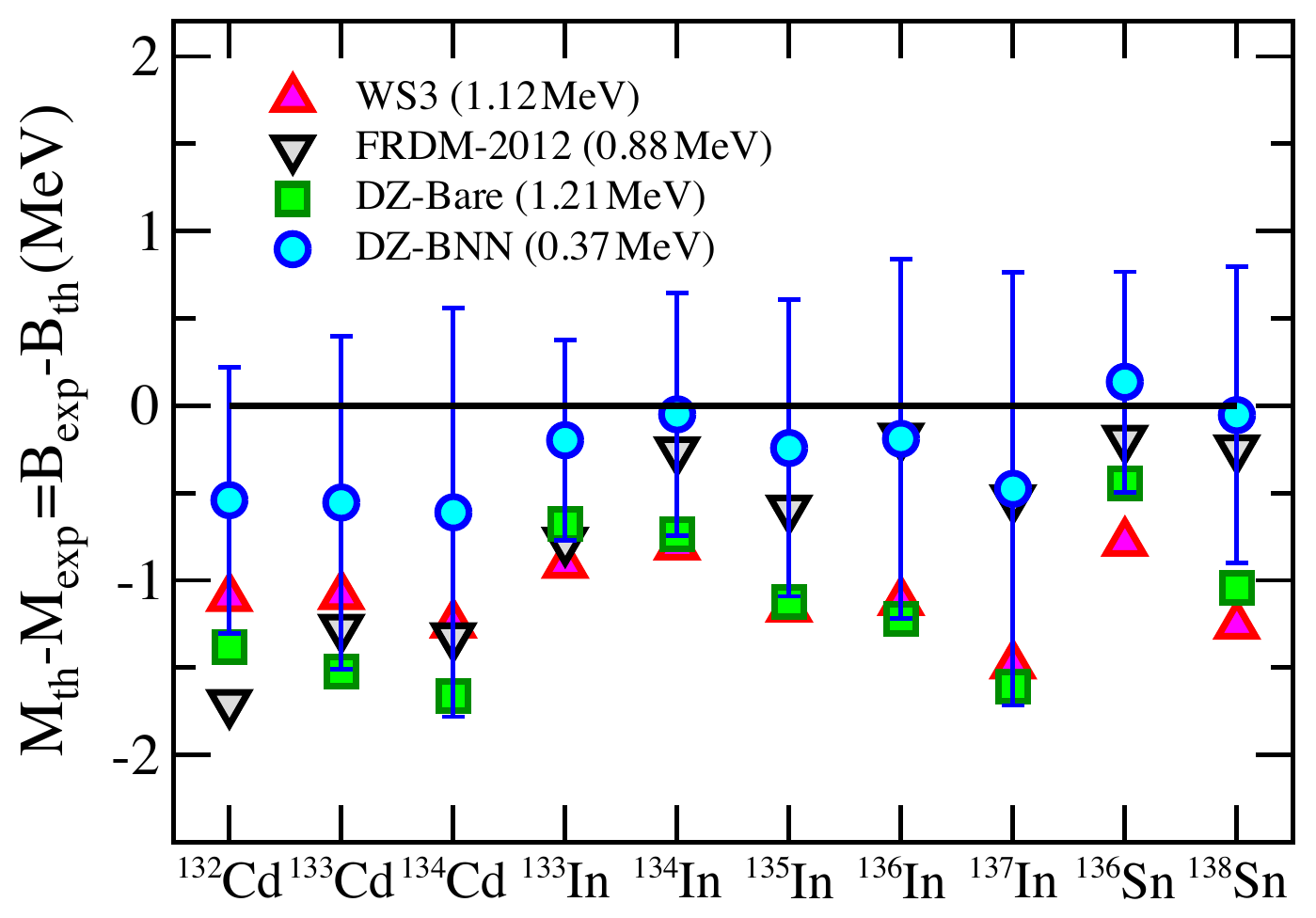}
\caption{Theoretical predictions for the total binding energy of those
nuclei that have been identified as impactful in $r$-process 
nucleosynthesis\,\cite{Mumpower:2015hva}. All experimental values 
have been estimated from experimental trends of neighboring 
nuclides\,\cite{AME:2016}. Quantities in parentheses denote the rms 
deviations.}
\label{Fig4}
\end{figure}

We close this section with a small comment on the impact of our work on 
the historical first detection of a binary neutron star merger by the LIGO-Virgo 
collaboration\,\cite{Abbott:PRL2017}, an event that is starting to provide 
fundamental new insights into the astrophysical site for the $r$-process 
and on the nature of dense matter. In particular, we focus on a particular 
set of nuclear masses that have been identified as ``impactful" in sensitivity 
studies of the elemental abundances in $r$-process nucleosynthesis. These 
include a variety of neutron-rich isotopes in cadmium, indium, and tin; see 
Table I of Ref.\,\cite{Mumpower:2015hva}. In Fig.\,\ref{Fig4} theoretical 
predictions are displayed for the mass of some of these isotopes. Predictions 
are provided for the WS3\,\cite{Liu:2011ama}, FRDM-2012\,\cite{Moller:2012}, 
DZ\,\cite{Duflo:1995}, and BNN-DZ\,\cite{Utama:2017wqe} mass models. 
Root-mean-square deviations of the order of 1\,MeV are recorded for all
models, except for the BNN-improved Duflo-Zuker model where the deviation
is only 369\,keV. The figure nicely encapsulates the spirit of our two-prong 
approach, namely, one that starts with a mass model of the highest quality 
(DZ) that is then refined via a BNN approach. The improvement in the 
description of the experimental data together with a proper assessment
of the theoretical uncertainties are two of the greatest virtues of the BNN 
approach. Indeed, the BNN-DZ predictions are consistent with all the
masses of the impactful nuclei displayed in the figure and that have been
recently reported in the latest AME2016 mass compilation\,\cite{AME:2016}.

\subsection{The Inner Crust: Coulomb Frustration and Nuclear Pasta}
\label{sec:innercrust}

Although not covered in the lectures because of lack of time, a few
comments on the fascinating physics of the inner crust are pertinent.
Note that there have been various significant contributions from the
Brazilian community to this topic, both from inside\,\cite{Avancini:2008zz,
Avancini:2008kg,Avancini:2010ch,Avancini:2012bj} and outside of 
Brazil\,\cite{Schneider:2013dwa,Caplan:2014gaa,Horowitz:2014xca}.

The inner stellar crust comprises the region from neutron-drip density 
up to the density at which uniformity in the system is restored; see 
Sec.\,\ref{sec:liquidcore}. Yet the transition from the highly-ordered 
crystal to the uniform liquid is both interesting and complex. This is
because distance scales that were well separated in both the
crystalline phase (where the long-range Coulomb interaction dominates)
and in the uniform phase (where the short-range strong interaction
dominates) become comparable. This unique situation gives rise to {\sl
``Coulomb frustration''}.  Frustration, a phenomenon characterized by
the existence of a very large number of low-energy configurations,
emerges from the impossibility to simultaneously minimize all
elementary interactions in the system. Indeed, as these length scales
become comparable, competition among the elementary interactions
results in the formation of a myriad of complex structures radically
different in topology yet extremely close in energy.  Given that these
complex structures---collectively referred to as \emph{``nuclear
pasta''}---are very close in energy, it has been speculated that the
transition from the highly ordered crystal to the uniform phase must
proceed through a series of changes in the dimensionality and topology
of these structures\,\cite{Ravenhall:1983uh,Hashimoto:1984}. Moreover,
due to the preponderance of low-energy states, frustrated systems
display an interesting and unique low-energy dynamics that has been
captured using a variety of techniques including semi-classical numerical
simulations\,\cite{Horowitz:2004yf,Horowitz:2004pv,
Horowitz:2005zb,Watanabe:2003xu,Watanabe:2004tr, Watanabe:2009vi,
Schneider:2013dwa,Horowitz:2014xca,Caplan:2014gaa} as well as
quantum simulations in a mean-field approximation\,\cite{Bulgac:2001,
Magierski:2001ud,Chamel:2004in,Newton:2009zz,Schuetrumpf:2015nza}.
For some extensive reviews on the fascinating structure and dynamics of 
the neutron-star crust see Refs.\,\cite{Chamel:2008ca,Bertulani:2012}, and 
references contain therein. 

\begin{figure}[h]
 \includegraphics[width=0.8\columnwidth]{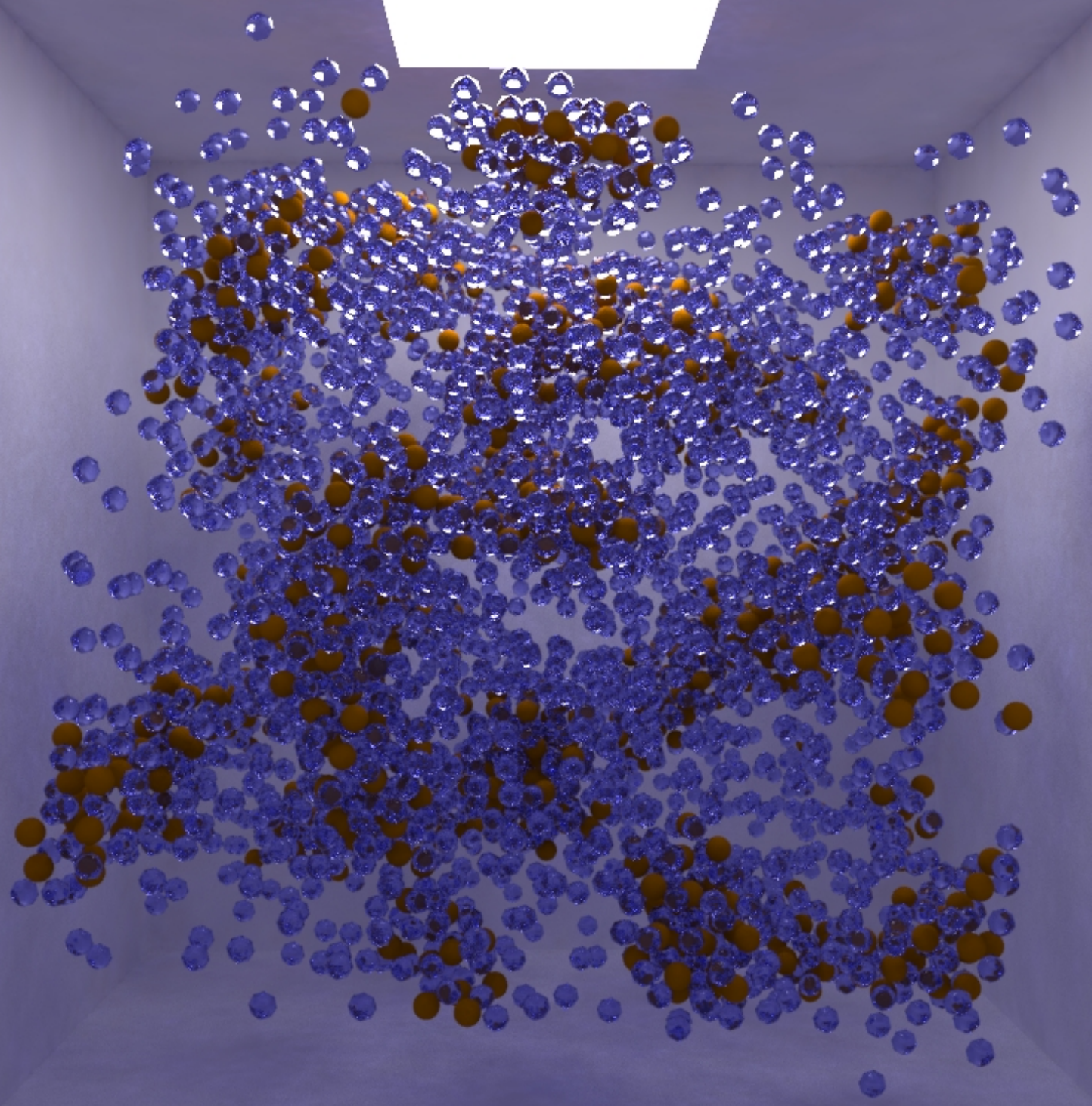}
 \caption{A snapshot of a Monte-Carlo simulation for a system consisting 
 of 4000 nucleons at a baryon density of  $\rho\!=\rhoz/6$, 
 a proton fraction of $Z/A\!=\!0.2$, and an ``effective" temperature of 
 $T\!=\!1$\,\,MeV\,\cite{Horowitz:2004yf,Horowitz:2004pv}.} 
 \label{Fig5}
\end{figure}

In closing this section, we display in Fig.\,\ref{Fig5} a Monte Carlo snapshot 
obtained from a numerical simulation of a system containing $Z\!=\!800$\,protons 
and $N\!=\!3200$\,neutrons that nicely illustrates how the system organizes itself 
into neutron-rich clusters of complex topologies that are immersed in a dilute 
neutron vapor\,\cite{Horowitz:2004yf,Horowitz:2004pv}. We note that a great 
virtue of these numerical simulations is that it clearly illustrates how pasta 
formation is very robust. Indeed, our numerical simulations proceed in an 
unbiased manner without assuming any particular shape. Instead, the 
system evolves dynamically into these complex shapes from a simple 
underlying two-body interaction consisting of a short-range nuclear attraction 
and a long-range Coulomb repulsion.

\subsection{The Liquid Core: Uniform Neutron-Rich Matter}
\label{sec:liquidcore}

At densities of about $10^{14}{\rm g/cm^{3}}$ the common perception of a 
neutron star as a uniform assembly of extremely closed packed neutrons
is finally realized in the stellar core. As we articulate below, the liquid core
is responsible for the most salient structural features of a neutron star, such
as its mass and its radius. Given the unique synergy between laboratory 
experiments and astrophysical observations, we devote the entire next 
section to illustrate these connections. However,
nowhere in these discussions we examine in detail the possibility of exotic
states of matter harboring the stellar core. Rather, we push our formalism 
consisting of conventional constituents (nucleons and charged 
leptons) to the extremes of density and neutron-proton asymmetry. 
Evidence in favor of exotic degrees of freedom may then emerge as our
accurately calibrated models show serious discrepancies when compared
against observations.

\section{Heaven and Earth}
\label{sec:heavenearth}

\begin{figure}[h]
\includegraphics[width=0.95\columnwidth]{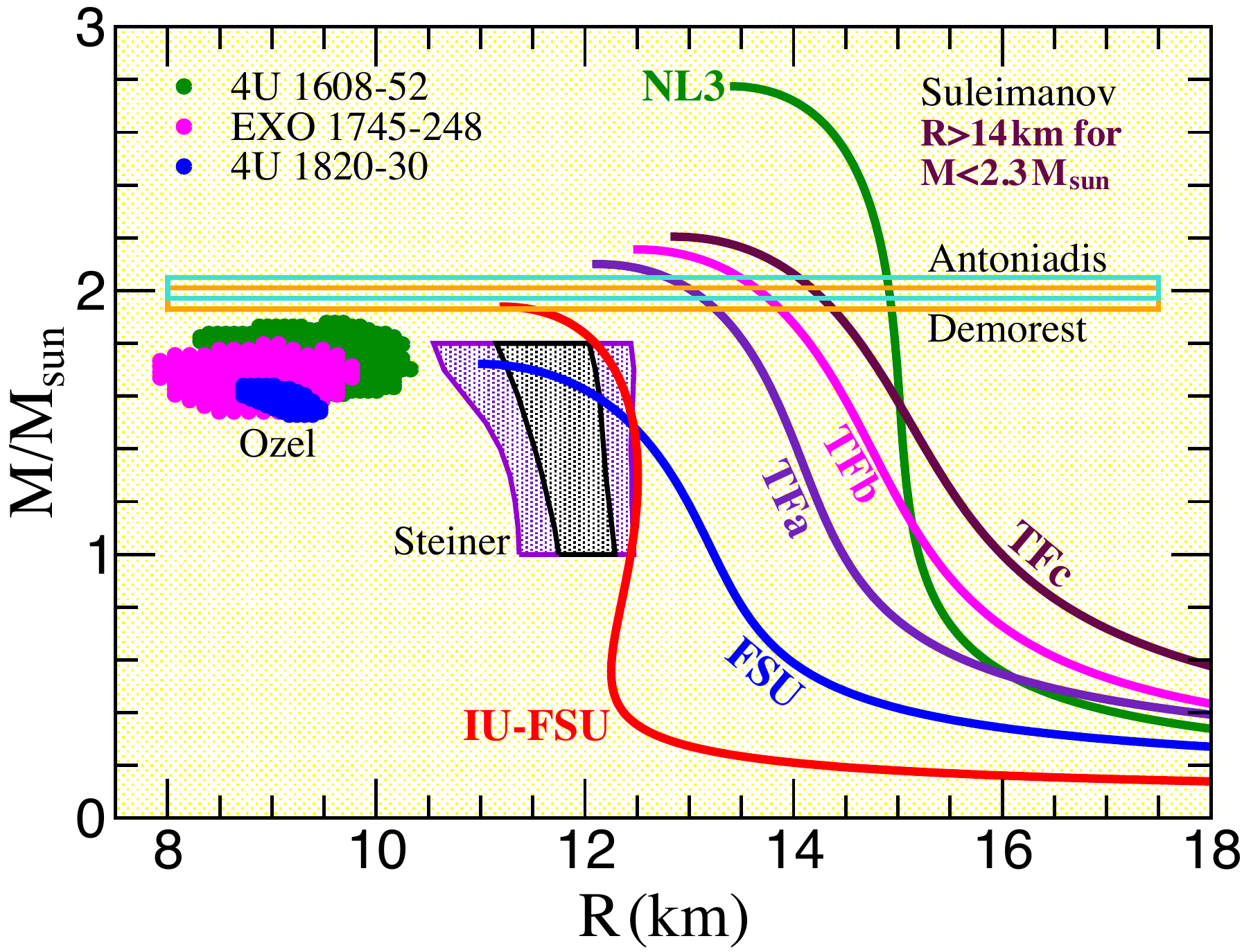}
 \vspace{-0.2cm}
   \caption{Predictions for the mass-vs-radius relation for a variety 
                 of relativistic models of the EOS\,\cite{Fattoyev:2013yaa}. 
                 Photometric constraints
                 on stellar masses and radii extracted from various analyses 
                 of X-ray bursts are shown\,\cite{Ozel:2010fw,Steiner:2010fz,
                 Suleimanov:2010th}. Also shown are constraints obtained 
                 from the measurement of two massive neutron stars by 
                 Demorest\,\cite{Demorest:2010bx} and 
                 Antoniadis\,\cite{Antoniadis:2013pzd}.}
 \label{Fig6}
\end{figure}

At densities of about a half to a third of nuclear-matter saturation density 
the nuclear-pasta phase will ``melt'' and uniformity in the system will be
restored. However, in order to maintain both chemical equilibrium and
charge neutrality a small fraction of about 10\% of protons and charged
leptons (electrons and muons) is required. Although the stellar crust is
driven by unique and intriguing dynamics, its structural impact on the 
star is rather modest. Indeed, more than 90\% of the size and essentially
all the stellar mass reside in the core. However, the equation of state 
of neutron-rich matter at the highest densities attained in the core is poorly
constrained by laboratory observables. Thus, the cleanest constraint on the 
EOS at high density is likely to emerge from astrophysical observations
of massive neutron stars. In this regard, enormous progress has been 
made with the observation of two massive neutron stars by 
Demorest\,\cite{Demorest:2010bx} and Antoniadis\,\cite{Antoniadis:2013pzd}. 
For example, the measurement of the 
mass of PSR J164-2230 $(1.97\!\pm\!0.04\,M_{\odot})$ by itself has ruled 
out EOS that are too soft to support a $2\,M_{\odot}$ neutron star---such
as those with exotic cores. Undoubtedly, the quest to find even more massive 
neutron stars will continue with the deployment of new missions, such as the 
\emph{Neutron Star Interior Composition Explorer} (NICER) and with the 
imminent detection of more binary neutron-star mergers. 
Predictions for the mass-vs-radius relation for a variety of relativistic models 
consistent with the $2M_{\odot}$ limit, with the exception of 
FSUGold\,\cite{Todd-Rutel:2005fa} (``FSU" in the figure) are displayed in 
Fig.\,\ref{Fig6}. We note that all our calculations have been done using models 
that yield an accurate description of the properties of finite nuclei, while 
providing a Lorentz covariant extrapolation to dense 
matter\,\cite{Fattoyev:2013yaa}. This implies that by construction, the EOS 
remains causal at all densities.

Unfortunately, the extraction of stellar radii by photometric means has been 
notoriously challenging, as it has been plagued by large systematic 
uncertainties, often revealing discrepancies as large as
5-6\,km\,\cite{Ozel:2010fw,Steiner:2010fz,Suleimanov:2010th}; see
Fig.\,\ref{Fig6}. It appears, however, that the situation has improved through 
a better understanding of systematic uncertainties, important theoretical 
developments, and the implementation of robust statistical 
methods\,\cite{Guillot:2013wu,Lattimer:2013hma,Heinke:2014xaa,Guillot:2014lla,
Ozel:2015fia,Watts:2016uzu,Steiner:2017vmg,Nattila:2017wtj}. As we will show 
later, stellar radii also leave their imprint on the gravitational wave form measured
in the merger of two neutron stars and will play a critical role as detections 
of these mergers become more plentiful.

Unlike massive neutron stars, stellar radii are sensitive to the density dependence 
of the symmetry energy in the immediate vicinity of nuclear-matter saturation 
density\,\cite{Lattimer:2006xb}. This may prove advantageous, as  the symmetry
energy at moderate densities may be constrained by terrestrial experiment. A 
fundamental property of the EOS that has received considerable attention over 
the last decade is the slope of the symmetry energy at saturation density.
The symmetry energy is an essential component of the equation of state that 
strongly impacts the structure, dynamics, and composition of neutron stars. 
Before chemical equilibrium is enforced, the equation of state depends on the 
conserved baryon density $\rho\!=\rho_{n}\!+\!\rho_{p}$ and the neutron-proton 
asymmetry $\alpha\!\equiv\!(\rho_{n}\!-\!\rho_{p})/(\rho_{n}\!+\!\rho_{p})$. As it is 
customarily done, the energy per nucleon may be expanded at zero 
temperature in even powers of $\alpha$:
\begin{equation}
  \frac{E}{A}(\rho,\alpha) -\!M \equiv {\cal E}(\rho,\alpha)
                          = {\cal E}_{\rm SNM}(\rho)
                          + \alpha^{2}{\cal S}(\rho)  
                          + {\cal O}(\alpha^{4}) \,,
 \label{EOS}
\end {equation}
where ${\cal E}_{\rm SNM}(\rho)\!=\!{\cal E}(\rho,\alpha\!\equiv\!0)$
is the energy per nucleon of \emph{symmetric} nuclear matter (SNM) 
and ${\cal S}(\rho)$ is the \emph{symmetry energy}, which represents 
the first-order correction to the symmetric limit. More intuitively, the
symmetry energy is nearly equal to the energy cost required to 
convert symmetric nuclear matter, with $\alpha\!=\!0$, into pure 
neutron matter (PNM) with $\alpha\!=\!1$:
\begin{equation}
 {\cal S}(\rho)\!\approx\!{\cal E}(\rho,\alpha\!=\!1) \!-\! 
 {\cal E}(\rho,\alpha\!=\!0) \;.
 \label{SymmE}
\end {equation}
Note that no odd powers of $\alpha$ appear in the expansion
as the nuclear force is assumed to be isospin symmetric and 
(for now) electroweak contributions have been ``turned off''. 
Finally, it is customary to characterize the behavior of both 
symmetric nuclear matter and the symmetry energy near 
saturation density in terms of a few bulk parameters. To do so, 
one performs a Taylor series expansion around nuclear matter 
saturation density $\rhoz$. That is\,\cite{Piekarewicz:2008nh},
\begin{subequations}
\begin{align}
 & {\cal E}_{\rm SNM}(\rho) = \epsz + \frac{1}{2}Kx^{2}+\ldots ,\label{EandSa}\\
 & {\cal S}(\rho) = J + Lx + \frac{1}{2}K_{\rm sym}x^{2}+\ldots ,\label{EandSb}
\end{align} 
\label{EandS}
\end{subequations}
\!\!\!where $x\!=\!(\rho-\rhoz)\!/3\rhoz$ is a dimensionless parameter that
quantifies the deviations of the density from its value at saturation. Here
$\epsz$ and $K$ represent the energy per nucleon and the 
incompressibility coefficient of SNM; $J$ and 
$K_{\rm sym}$ are the corresponding quantities for the symmetry energy. 
However, unlike symmetric nuclear matter whose pressure vanishes at
$\rhoz$, the slope of the symmetry energy $L$ does not vanish at
saturation density. Indeed, assuming the validity of Eq.\,(\ref{SymmE}),
$L$ is directly proportional to the pressure of PNM ($P_{0}$) at saturation 
density, namely,
\begin{equation}
   P_{0} \approx \frac{1}{3}\rhoz L \;.
 \label{PvsL}
\end{equation}

Given that neutron-star radii are sensitive to the density dependence of 
the \emph{symmetry energy} in the vicinity of nuclear-matter saturation
density\,\cite{Lattimer:2006xb}, laboratory experiments play a critical
role in constraining stellar radii.  Indeed, $L$ is strongly correlated to 
both the thickness of the neutron skin in ${}^{208}$Pb\,\cite{Brown:2000,
Furnstahl:2001un,Centelles:2008vu,RocaMaza:2011pm} and the radius 
of a neutron star\,\cite{Horowitz:2000xj,Horowitz:2001ya,Carriere:2002bx,
Fattoyev:2012rm,Chen:2014sca}. Note that the neutron-skin thickness is 
defined as the difference between the neutron and proton root-mean-square 
radii. The thickness of the neutron skin emerges from a competition between 
the surface tension, which favors placing the excess neutrons in the interior,
and the \emph{difference} between the value of the symmetry energy at 
the surface relative to that at the center; namely, $L$. If such a difference 
is large, then it is favorable to move the extra neutrons to the surface, 
thereby creating a thick neutron skin. Similarly, if the pressure of pure
neutron matter at saturation is large, a large stellar radius develops.
This suggests a powerful correlation: \emph{the larger the value of $L$ 
the thicker the neutron skin and the larger the radius of a neutron 
star}\,\cite{Horowitz:2001ya}. In this way, the neutron-skin thickness 
in ${}^{208}$Pb is identified as a laboratory observable that may serve 
to constrain the radius of a neutron star---despite a difference in 
size of 19 orders of magnitude! 

Using a purely electroweak reaction---parity-violating electron 
scattering---the pioneering Lead Radius Experiment (``PREX") 
at the Jefferson Laboratory provided the first model-independent 
evidence in favor of a neutron-rich skin in
${}^{208}$Pb\,\cite{Abrahamyan:2012gp,Horowitz:2012tj}: 
$R_{\rm skin}^{208}\!=\!{0.33}^{+0.16}_{-0.18}\,{\rm fm}.$ Unfortunately, 
the larger than anticipated statistical error has hindered a meaningful
comparison against theoretical predictions. Yet, in an effort to impose 
meaningful theoretical constraints, a follow-up experiment (PREX-II) 
is scheduled to run in 2019 that is envisioned to reach the original
$0.06$\,fm sensitivity.

Finally, constraints on the density dependence of the symmetry energy
have emerged from an unexpected source: the historical first detection 
of the binary neutron-star merger GW170817\,\cite{Abbott:PRL2017}. 
The \emph{tidal polarizability} of a neutron star, namely, the tendency 
to develop a mass quadrupole as a response to the tidal field induced 
by its companion\,\cite{Damour:1991yw,Flanagan:2007ix}, is imprinted 
in the gravitational wave form associated with the binary inspiral.
Indeed, the gravitational wave form maintains its point-mass 
(black-hole-like) behavior longer for compact stars than for stars with 
larger radii. In a recent publication that examined the impact of GW170817
on the tidal polarizability, we inferred a limit on the stellar radius of a
$1.4\,M_{\odot}$ neutron star of 
$R_{\star}^{1.4}\!<\!13.76\,{\rm km}$\,\cite{Fattoyev:2017jql}. In the
context of Fig.\,\ref{Fig6}, this is a highly significant result. With the
exception of IU-FSU\,\cite{Fattoyev:2010mx}, all models displayed
in the figure are ruled out either because they predict a maximum 
mass that is too low or a stellar radius that is too large. Further, assuming 
that one can extrapolate our findings down to saturation density, 
constraints from GW170817 also provide limits on the neutron-skin 
thickness of ${}^{208}$Pb of $R_{\rm skin}^{208}\!\lesssim\!0.25\,{\rm fm}$, 
well below the current upper limit obtained by the PREX 
collaboration\,\cite{Abrahamyan:2012gp,Horowitz:2012tj}. This suggest
an intriguing possibility: If PREX-II confirms that 
$R_{\rm skin}^{208}$ is large, this will suggest that the EOS at the 
typical densities found in atomic nuclei is stiff. In contrast, the 
relatively small neutron-star radii suggested by GW170817 implies 
that the symmetry energy at higher densities is soft. The evolution 
from stiff to soft may be indicative of a phase transition in the 
neutron-star interior. 

\section{Conclusions and Outlook}
\label{sec:conclusions}

Neutron stars are gold mines for the study of physical phenomena across 
a variety of disciplines ranging from the very small to the very large, from
elementary-particle physics to general relativity. From the perspective of 
hadronic and nuclear physics, the main topic of the XIV International 
Workshop, neutron stars hold the answer to one of the most fundamental 
questions in the field: \emph{How does subatomic matter organize itself 
and what phenomena emerge?}\,\cite{national2012Nuclear}. Although 
the most common perception of a neutron star is a uniform ensemble of 
neutrons, we showed that the reality is far different and much more 
interesting. In particular, during our journey through a neutron star we 
uncovered a myriad of exotic states of matter that are speculated to 
exist in a neutron star, such as Coulomb crystals, pasta phases, and 
perhaps even deconfined quark matter. As exciting, we discussed the 
fundamental role that nuclear astrophysics will play in the new era of 
multimessenger astronomy. Although binary pulsars---such as the 
Hulse-Taylor pulsar---have been used to infer the existence of gravitational 
waves, the evidence was indirect. Now, however, we have the first direct 
evidence of gravitational waves from a binary neutron-star merger. In a
testament to human ingenuity, many of the observed phenomena 
associated with the binary neutron-star merger were predicted by 
earlier theoretical simulations. Surprisingly, this very first observation 
has provided a treasure trove of insights into the nature of dense matter
and the site of the $r$-process. 

Yet the era of multimessenger astronomy is in its infancy and much 
excitement is in store. Electromagnetic, gravitational and, hopefully
soon, neutrino radiation from spectacular neutron-star mergers will 
reveal some of nature's most intimate secrets. This new era is of 
particular significance for the Hadron Physics series whose summer 
school format provides an ideal venue to educate and motivate the 
next generation of scientists. Undoubtedly, nuclear and hadronic 
physics will play a fundamental role in elucidating the physics 
underlying these spectacular events. And it is the new generation
of scientists that will reap the benefits from this scientific revolution
and who will make the new discoveries. I hope that through this set 
of lectures that I was privileged to deliver, I was able to inspire many 
young scientists to join this fascinating field.

\begin{acknowledgments}
 \vspace{-0.4cm}
 I am very grateful to the organizers of the XIV International Workshop
 in Hadron Physics, particularly Prof. Menezes, Prof. Benghi, and 
 Dr. Oliveira, for their kindness and hospitality. The financial support of
 the CNPq is greatly appreciated. This material is 
 based upon work supported by the U.S. Department of Energy Office 
 of Science, Office of Nuclear Physics under Award Number 
 DE-FG02-92ER40750.
\end{acknowledgments}

\bibliography{Proceedings.bbl}

\begin{thebibliography}{90}
\expandafter\ifx\csname natexlab\endcsname\relax\def\natexlab#1{#1}\fi
\expandafter\ifx\csname bibnamefont\endcsname\relax
  \def\bibnamefont#1{#1}\fi
\expandafter\ifx\csname bibfnamefont\endcsname\relax
  \def\bibfnamefont#1{#1}\fi
\expandafter\ifx\csname citenamefont\endcsname\relax
  \def\citenamefont#1{#1}\fi
\expandafter\ifx\csname url\endcsname\relax
  \def\url#1{\texttt{#1}}\fi
\expandafter\ifx\csname urlprefix\endcsname\relax\def\urlprefix{URL }\fi
\providecommand{\bibinfo}[2]{#2}
\providecommand{\eprint}[2][]{\url{#2}}

\bibitem[{\citenamefont{Abbott et~al.}(2017)}]{Abbott:PRL2017}
\bibinfo{author}{\bibfnamefont{B.~P.} \bibnamefont{Abbott}}
  \bibnamefont{et~al.} (\bibinfo{collaboration}{Virgo, LIGO Scientific}),
  \bibinfo{journal}{Phys. Rev. Lett.} \textbf{\bibinfo{volume}{119}},
  \bibinfo{pages}{161101} (\bibinfo{year}{2017}).

\bibitem[{Qua(2003)}]{QuarksCosmos:2003}
\emph{\bibinfo{title}{Connecting Quarks with the Cosmos: Eleven Science
  Questions for the New Century}} (\bibinfo{publisher}{The National Academies
  Press}, \bibinfo{address}{Washington}, \bibinfo{year}{2003}).

\bibitem[{\citenamefont{Oppenheimer and Volkoff}(1939)}]{Opp39_PR55}
\bibinfo{author}{\bibfnamefont{J.~R.} \bibnamefont{Oppenheimer}}
  \bibnamefont{and} \bibinfo{author}{\bibfnamefont{G.~M.}
  \bibnamefont{Volkoff}}, \bibinfo{journal}{Phys. Rev.}
  \textbf{\bibinfo{volume}{55}}, \bibinfo{pages}{374} (\bibinfo{year}{1939}).

\bibitem[{\citenamefont{Chadwick}(1932)}]{Chadwick:1932}
\bibinfo{author}{\bibfnamefont{J.}~\bibnamefont{Chadwick}},
  \bibinfo{journal}{Nature} \textbf{\bibinfo{volume}{129}},
  \bibinfo{pages}{312} (\bibinfo{year}{1932}).

\bibitem[{\citenamefont{Baade and Zwicky}(1934)}]{Baade:1934}
\bibinfo{author}{\bibfnamefont{W.}~\bibnamefont{Baade}} \bibnamefont{and}
  \bibinfo{author}{\bibfnamefont{F.}~\bibnamefont{Zwicky}},
  \bibinfo{journal}{Phys. Rev.} \textbf{\bibinfo{volume}{45}},
  \bibinfo{pages}{138} (\bibinfo{year}{1934}).

\bibitem[{\citenamefont{Demorest et~al.}(2010)\citenamefont{Demorest, Pennucci,
  Ransom, Roberts, and Hessels}}]{Demorest:2010bx}
\bibinfo{author}{\bibfnamefont{P.}~\bibnamefont{Demorest}},
  \bibinfo{author}{\bibfnamefont{T.}~\bibnamefont{Pennucci}},
  \bibinfo{author}{\bibfnamefont{S.}~\bibnamefont{Ransom}},
  \bibinfo{author}{\bibfnamefont{M.}~\bibnamefont{Roberts}}, \bibnamefont{and}
  \bibinfo{author}{\bibfnamefont{J.}~\bibnamefont{Hessels}},
  \bibinfo{journal}{Nature} \textbf{\bibinfo{volume}{467}},
  \bibinfo{pages}{1081} (\bibinfo{year}{2010}).

\bibitem[{\citenamefont{Antoniadis et~al.}(2013)\citenamefont{Antoniadis,
  Freire, Wex, Tauris, Lynch et~al.}}]{Antoniadis:2013pzd}
\bibinfo{author}{\bibfnamefont{J.}~\bibnamefont{Antoniadis}},
  \bibinfo{author}{\bibfnamefont{P.~C.} \bibnamefont{Freire}},
  \bibinfo{author}{\bibfnamefont{N.}~\bibnamefont{Wex}},
  \bibinfo{author}{\bibfnamefont{T.~M.} \bibnamefont{Tauris}},
  \bibinfo{author}{\bibfnamefont{R.~S.} \bibnamefont{Lynch}},
  \bibnamefont{et~al.}, \bibinfo{journal}{Science}
  \textbf{\bibinfo{volume}{340}}, \bibinfo{pages}{6131} (\bibinfo{year}{2013}).

\bibitem[{\citenamefont{Chandrasekhar}(1931)}]{Chandrasekhar:1931}
\bibinfo{author}{\bibfnamefont{S.}~\bibnamefont{Chandrasekhar}},
  \bibinfo{journal}{Astrophys. J} \textbf{\bibinfo{volume}{74}},
  \bibinfo{pages}{81} (\bibinfo{year}{1931}).

\bibitem[{\citenamefont{Landau}(1932)}]{Landau:1932}
\bibinfo{author}{\bibfnamefont{L.~D.} \bibnamefont{Landau}},
  \bibinfo{journal}{Phys. Z. Sowjetunion} \textbf{\bibinfo{volume}{1}},
  \bibinfo{pages}{285} (\bibinfo{year}{1932}), \bibinfo{note}{translated into
  Russian: in Landau L D Sobranie Trudov (Collected Works) Vol. 1 (Moscow:
  Nauka, 1969) p. 86}.

\bibitem[{\citenamefont{Yakovlev et~al.}(2013)\citenamefont{Yakovlev, Haensel,
  Baym, and Pethick}}]{Yakovlev:2012rd}
\bibinfo{author}{\bibfnamefont{D.~G.} \bibnamefont{Yakovlev}},
  \bibinfo{author}{\bibfnamefont{P.}~\bibnamefont{Haensel}},
  \bibinfo{author}{\bibfnamefont{G.}~\bibnamefont{Baym}}, \bibnamefont{and}
  \bibinfo{author}{\bibfnamefont{C.~J.} \bibnamefont{Pethick}},
  \bibinfo{journal}{Phys. Usp.} \textbf{\bibinfo{volume}{56}},
  \bibinfo{pages}{289} (\bibinfo{year}{2013}), \eprint{1210.0682}.

\bibitem[{\citenamefont{Meng}(2016)}]{Meng:2016}
\bibinfo{author}{\bibfnamefont{J.}~\bibnamefont{Meng}},
  \emph{\bibinfo{title}{Relativistic Density Functional for Nuclear Structure}}
  (\bibinfo{publisher}{World Scientific}, \bibinfo{address}{New Jersey},
  \bibinfo{year}{2016}), International Review of Nuclear Physics: Volume 10.

\bibitem[{\citenamefont{Hewish et~al.}(1968)\citenamefont{Hewish, Bell,
  Pilkington, Scott, and Collins}}]{Hewish:1968}
\bibinfo{author}{\bibfnamefont{A.}~\bibnamefont{Hewish}},
  \bibinfo{author}{\bibfnamefont{S.}~\bibnamefont{Bell}},
  \bibinfo{author}{\bibfnamefont{J.}~\bibnamefont{Pilkington}},
  \bibinfo{author}{\bibfnamefont{P.}~\bibnamefont{Scott}}, \bibnamefont{and}
  \bibinfo{author}{\bibfnamefont{R.}~\bibnamefont{Collins}},
  \bibinfo{journal}{Nature} \textbf{\bibinfo{volume}{217}},
  \bibinfo{pages}{709} (\bibinfo{year}{1968}).

\bibitem[{\citenamefont{Page et~al.}(2004)\citenamefont{Page, Lattimer,
  Prakash, and Steiner}}]{Page:2004fy}
\bibinfo{author}{\bibfnamefont{D.}~\bibnamefont{Page}},
  \bibinfo{author}{\bibfnamefont{J.~M.} \bibnamefont{Lattimer}},
  \bibinfo{author}{\bibfnamefont{M.}~\bibnamefont{Prakash}}, \bibnamefont{and}
  \bibinfo{author}{\bibfnamefont{A.~W.} \bibnamefont{Steiner}},
  \bibinfo{journal}{Astrophys. J. Suppl.} \textbf{\bibinfo{volume}{155}},
  \bibinfo{pages}{623} (\bibinfo{year}{2004}).

\bibitem[{\citenamefont{Ellis et~al.}(1995)\citenamefont{Ellis, Knorren, and
  Prakash}}]{Ellis:1995kz}
\bibinfo{author}{\bibfnamefont{P.~J.} \bibnamefont{Ellis}},
  \bibinfo{author}{\bibfnamefont{R.}~\bibnamefont{Knorren}}, \bibnamefont{and}
  \bibinfo{author}{\bibfnamefont{M.}~\bibnamefont{Prakash}},
  \bibinfo{journal}{Phys. Lett.} \textbf{\bibinfo{volume}{B349}},
  \bibinfo{pages}{11} (\bibinfo{year}{1995}).

\bibitem[{\citenamefont{Pons et~al.}(2001)\citenamefont{Pons, Miralles,
  Prakash, and Lattimer}}]{Pons:2000xf}
\bibinfo{author}{\bibfnamefont{J.~A.} \bibnamefont{Pons}},
  \bibinfo{author}{\bibfnamefont{J.~A.} \bibnamefont{Miralles}},
  \bibinfo{author}{\bibfnamefont{M.}~\bibnamefont{Prakash}}, \bibnamefont{and}
  \bibinfo{author}{\bibfnamefont{J.~M.} \bibnamefont{Lattimer}},
  \bibinfo{journal}{Astrophys. J.} \textbf{\bibinfo{volume}{553}},
  \bibinfo{pages}{382} (\bibinfo{year}{2001}).

\bibitem[{\citenamefont{Weber}(2005)}]{Weber:2004kj}
\bibinfo{author}{\bibfnamefont{F.}~\bibnamefont{Weber}},
  \bibinfo{journal}{Prog. Part. Nucl. Phys.} \textbf{\bibinfo{volume}{54}},
  \bibinfo{pages}{193} (\bibinfo{year}{2005}).

\bibitem[{\citenamefont{Page and Reddy}(2006)}]{Page:2006ud}
\bibinfo{author}{\bibfnamefont{D.}~\bibnamefont{Page}} \bibnamefont{and}
  \bibinfo{author}{\bibfnamefont{S.}~\bibnamefont{Reddy}},
  \bibinfo{journal}{Ann. Rev. Nucl. Part. Sci.} \textbf{\bibinfo{volume}{56}},
  \bibinfo{pages}{327} (\bibinfo{year}{2006}).

\bibitem[{\citenamefont{Alford et~al.}(1999)\citenamefont{Alford, Rajagopal,
  and Wilczek}}]{Alford:1998mk}
\bibinfo{author}{\bibfnamefont{M.~G.} \bibnamefont{Alford}},
  \bibinfo{author}{\bibfnamefont{K.}~\bibnamefont{Rajagopal}},
  \bibnamefont{and} \bibinfo{author}{\bibfnamefont{F.}~\bibnamefont{Wilczek}},
  \bibinfo{journal}{Nucl. Phys.} \textbf{\bibinfo{volume}{B537}}
  (\bibinfo{year}{1999}).

\bibitem[{\citenamefont{Alford et~al.}(2008)\citenamefont{Alford, Schmitt,
  Rajagopal, and Schafer}}]{Alford:2007xm}
\bibinfo{author}{\bibfnamefont{M.~G.} \bibnamefont{Alford}},
  \bibinfo{author}{\bibfnamefont{A.}~\bibnamefont{Schmitt}},
  \bibinfo{author}{\bibfnamefont{K.}~\bibnamefont{Rajagopal}},
  \bibnamefont{and} \bibinfo{author}{\bibfnamefont{T.}~\bibnamefont{Schafer}},
  \bibinfo{journal}{Rev. Mod. Phys.} \textbf{\bibinfo{volume}{80}},
  \bibinfo{pages}{1455} (\bibinfo{year}{2008}).

\bibitem[{\citenamefont{{Haensel} et~al.}(1989)\citenamefont{{Haensel},
  {Zdunik}, and {Dobaczewski}}}]{Haensel:1989}
\bibinfo{author}{\bibfnamefont{P.}~\bibnamefont{{Haensel}}},
  \bibinfo{author}{\bibfnamefont{J.~L.} \bibnamefont{{Zdunik}}},
  \bibnamefont{and}
  \bibinfo{author}{\bibfnamefont{J.}~\bibnamefont{{Dobaczewski}}},
  \bibinfo{journal}{Astron. Astrophys.} \textbf{\bibinfo{volume}{222}},
  \bibinfo{pages}{353} (\bibinfo{year}{1989}).

\bibitem[{\citenamefont{Haensel and Pichon}(1994)}]{Haensel:1993zw}
\bibinfo{author}{\bibfnamefont{P.}~\bibnamefont{Haensel}} \bibnamefont{and}
  \bibinfo{author}{\bibfnamefont{B.}~\bibnamefont{Pichon}},
  \bibinfo{journal}{Astron. Astrophys.} \textbf{\bibinfo{volume}{283}},
  \bibinfo{pages}{313} (\bibinfo{year}{1994}).

\bibitem[{\citenamefont{Ruester et~al.}(2006)\citenamefont{Ruester, Hempel, and
  Schaffner-Bielich}}]{Ruester:2005fm}
\bibinfo{author}{\bibfnamefont{S.~B.} \bibnamefont{Ruester}},
  \bibinfo{author}{\bibfnamefont{M.}~\bibnamefont{Hempel}}, \bibnamefont{and}
  \bibinfo{author}{\bibfnamefont{J.}~\bibnamefont{Schaffner-Bielich}},
  \bibinfo{journal}{Phys. Rev.} \textbf{\bibinfo{volume}{C73}},
  \bibinfo{pages}{035804} (\bibinfo{year}{2006}).

\bibitem[{\citenamefont{Roca-Maza and Piekarewicz}(2008)}]{RocaMaza:2008ja}
\bibinfo{author}{\bibfnamefont{X.}~\bibnamefont{Roca-Maza}} \bibnamefont{and}
  \bibinfo{author}{\bibfnamefont{J.}~\bibnamefont{Piekarewicz}},
  \bibinfo{journal}{Phys. Rev.} \textbf{\bibinfo{volume}{C78}},
  \bibinfo{pages}{025807} (\bibinfo{year}{2008}).

\bibitem[{\citenamefont{Roca-Maza
  et~al.}(2011{\natexlab{a}})\citenamefont{Roca-Maza, Piekarewicz,
  Garcia-Galvez, and Centelles}}]{RocaMaza:2011pk}
\bibinfo{author}{\bibfnamefont{X.}~\bibnamefont{Roca-Maza}},
  \bibinfo{author}{\bibfnamefont{J.}~\bibnamefont{Piekarewicz}},
  \bibinfo{author}{\bibfnamefont{T.}~\bibnamefont{Garcia-Galvez}},
  \bibnamefont{and}
  \bibinfo{author}{\bibfnamefont{M.}~\bibnamefont{Centelles}}, in
  \emph{\bibinfo{booktitle}{Neutron Star Crust}}, edited by
  \bibinfo{editor}{\bibfnamefont{C.}~\bibnamefont{Bertulani}} \bibnamefont{and}
  \bibinfo{editor}{\bibfnamefont{J.}~\bibnamefont{Piekarewicz}}
  (\bibinfo{publisher}{Nova Publishers}, \bibinfo{address}{New York},
  \bibinfo{year}{2011}{\natexlab{a}}).

\bibitem[{\citenamefont{Wang et~al.}(2012)\citenamefont{Wang, Audi, Wapstra,
  Kondev, MacCormick, Xu, and Pfeiffer}}]{AME:2012}
\bibinfo{author}{\bibfnamefont{M.}~\bibnamefont{Wang}},
  \bibinfo{author}{\bibfnamefont{G.}~\bibnamefont{Audi}},
  \bibinfo{author}{\bibfnamefont{A.}~\bibnamefont{Wapstra}},
  \bibinfo{author}{\bibfnamefont{F.}~\bibnamefont{Kondev}},
  \bibinfo{author}{\bibfnamefont{M.}~\bibnamefont{MacCormick}},
  \bibinfo{author}{\bibfnamefont{X.}~\bibnamefont{Xu}}, \bibnamefont{and}
  \bibinfo{author}{\bibfnamefont{B.}~\bibnamefont{Pfeiffer}},
  \bibinfo{journal}{Chinese Phys. C} \textbf{\bibinfo{volume}{36}},
  \bibinfo{pages}{1603} (\bibinfo{year}{2012}).

\bibitem[{\citenamefont{Huang et~al.}(2017)\citenamefont{Huang, Audi, Wang,
  Kondev, Naimi, and Xu}}]{AME:2016}
\bibinfo{author}{\bibfnamefont{W.~J.} \bibnamefont{Huang}},
  \bibinfo{author}{\bibfnamefont{G.}~\bibnamefont{Audi}},
  \bibinfo{author}{\bibfnamefont{M.}~\bibnamefont{Wang}},
  \bibinfo{author}{\bibfnamefont{F.~G.} \bibnamefont{Kondev}},
  \bibinfo{author}{\bibfnamefont{S.}~\bibnamefont{Naimi}}, \bibnamefont{and}
  \bibinfo{author}{\bibfnamefont{X.}~\bibnamefont{Xu}}, \bibinfo{journal}{Chin.
  Phys.} \textbf{\bibinfo{volume}{C41}}, \bibinfo{pages}{030002}
  (\bibinfo{year}{2017}).

\bibitem[{\citenamefont{Utama et~al.}(2016{\natexlab{a}})\citenamefont{Utama,
  Piekarewicz, and Prosper}}]{Utama:2015hva}
\bibinfo{author}{\bibfnamefont{R.}~\bibnamefont{Utama}},
  \bibinfo{author}{\bibfnamefont{J.}~\bibnamefont{Piekarewicz}},
  \bibnamefont{and} \bibinfo{author}{\bibfnamefont{H.~B.}
  \bibnamefont{Prosper}}, \bibinfo{journal}{Phys. Rev.}
  \textbf{\bibinfo{volume}{C93}}, \bibinfo{pages}{014311}
  (\bibinfo{year}{2016}{\natexlab{a}}).

\bibitem[{\citenamefont{Utama et~al.}(2016{\natexlab{b}})\citenamefont{Utama,
  Chen, and Piekarewicz}}]{Utama:2016rad}
\bibinfo{author}{\bibfnamefont{R.}~\bibnamefont{Utama}},
  \bibinfo{author}{\bibfnamefont{W.-C.} \bibnamefont{Chen}}, \bibnamefont{and}
  \bibinfo{author}{\bibfnamefont{J.}~\bibnamefont{Piekarewicz}},
  \bibinfo{journal}{J. Phys.} \textbf{\bibinfo{volume}{G}}
  (\bibinfo{year}{2016}{\natexlab{b}}).

\bibitem[{\citenamefont{Utama and Piekarewicz}(2017)}]{Utama:2017wqe}
\bibinfo{author}{\bibfnamefont{R.}~\bibnamefont{Utama}} \bibnamefont{and}
  \bibinfo{author}{\bibfnamefont{J.}~\bibnamefont{Piekarewicz}},
  \bibinfo{journal}{Phys. Rev.} \textbf{\bibinfo{volume}{C96}},
  \bibinfo{pages}{044308} (\bibinfo{year}{2017}).

\bibitem[{\citenamefont{Utama and Piekarewicz}(2018)}]{Utama:2017ytc}
\bibinfo{author}{\bibfnamefont{R.}~\bibnamefont{Utama}} \bibnamefont{and}
  \bibinfo{author}{\bibfnamefont{J.}~\bibnamefont{Piekarewicz}},
  \bibinfo{journal}{Phys. Rev.} \textbf{\bibinfo{volume}{C97}},
  \bibinfo{pages}{014306} (\bibinfo{year}{2018}).

\bibitem[{\citenamefont{Piro}(2005)}]{Piro:2005jf}
\bibinfo{author}{\bibfnamefont{A.~L.} \bibnamefont{Piro}},
  \bibinfo{journal}{Astrophys. J.} \textbf{\bibinfo{volume}{634}},
  \bibinfo{pages}{L153} (\bibinfo{year}{2005}).

\bibitem[{\citenamefont{Steiner and Watts}(2009)}]{Steiner:2009yg}
\bibinfo{author}{\bibfnamefont{A.~W.} \bibnamefont{Steiner}} \bibnamefont{and}
  \bibinfo{author}{\bibfnamefont{A.~L.} \bibnamefont{Watts}},
  \bibinfo{journal}{Phys. Rev. Lett.} \textbf{\bibinfo{volume}{103}},
  \bibinfo{pages}{181101} (\bibinfo{year}{2009}).

\bibitem[{\citenamefont{Horowitz and Kadau}(2009)}]{Horowitz:2009ya}
\bibinfo{author}{\bibfnamefont{C.}~\bibnamefont{Horowitz}} \bibnamefont{and}
  \bibinfo{author}{\bibfnamefont{K.}~\bibnamefont{Kadau}},
  \bibinfo{journal}{Phys. Rev. Lett.} \textbf{\bibinfo{volume}{102}},
  \bibinfo{pages}{191102} (\bibinfo{year}{2009}).

\bibitem[{\citenamefont{Mumpower et~al.}(2015)\citenamefont{Mumpower, Surman,
  Fang, Beard, Moller, Kawano, and Aprahamian}}]{Mumpower:2015hva}
\bibinfo{author}{\bibfnamefont{M.~R.} \bibnamefont{Mumpower}},
  \bibinfo{author}{\bibfnamefont{R.}~\bibnamefont{Surman}},
  \bibinfo{author}{\bibfnamefont{D.~L.} \bibnamefont{Fang}},
  \bibinfo{author}{\bibfnamefont{M.}~\bibnamefont{Beard}},
  \bibinfo{author}{\bibfnamefont{P.}~\bibnamefont{Moller}},
  \bibinfo{author}{\bibfnamefont{T.}~\bibnamefont{Kawano}}, \bibnamefont{and}
  \bibinfo{author}{\bibfnamefont{A.}~\bibnamefont{Aprahamian}},
  \bibinfo{journal}{Phys. Rev.} \textbf{\bibinfo{volume}{C92}},
  \bibinfo{pages}{035807} (\bibinfo{year}{2015}).

\bibitem[{\citenamefont{Liu et~al.}(2011)\citenamefont{Liu, Wang, Deng, and
  Wu}}]{Liu:2011ama}
\bibinfo{author}{\bibfnamefont{M.}~\bibnamefont{Liu}},
  \bibinfo{author}{\bibfnamefont{N.}~\bibnamefont{Wang}},
  \bibinfo{author}{\bibfnamefont{Y.}~\bibnamefont{Deng}}, \bibnamefont{and}
  \bibinfo{author}{\bibfnamefont{X.}~\bibnamefont{Wu}}, \bibinfo{journal}{Phys.
  Rev.} \textbf{\bibinfo{volume}{C84}}, \bibinfo{pages}{014333}
  (\bibinfo{year}{2011}).

\bibitem[{\citenamefont{M\"oller et~al.}(2012)\citenamefont{M\"oller, Myers,
  Sagawa, and Yoshida}}]{Moller:2012}
\bibinfo{author}{\bibfnamefont{P.}~\bibnamefont{M\"oller}},
  \bibinfo{author}{\bibfnamefont{W.~D.} \bibnamefont{Myers}},
  \bibinfo{author}{\bibfnamefont{H.}~\bibnamefont{Sagawa}}, \bibnamefont{and}
  \bibinfo{author}{\bibfnamefont{S.}~\bibnamefont{Yoshida}},
  \bibinfo{journal}{Phys. Rev. Lett.} \textbf{\bibinfo{volume}{108}},
  \bibinfo{pages}{052501} (\bibinfo{year}{2012}).

\bibitem[{\citenamefont{Duflo and Zuker}(1995)}]{Duflo:1995}
\bibinfo{author}{\bibfnamefont{J.}~\bibnamefont{Duflo}} \bibnamefont{and}
  \bibinfo{author}{\bibfnamefont{A.}~\bibnamefont{Zuker}},
  \bibinfo{journal}{Phys. Rev.} \textbf{\bibinfo{volume}{C52}},
  \bibinfo{pages}{R23} (\bibinfo{year}{1995}).

\bibitem[{\citenamefont{Avancini et~al.}(2008)\citenamefont{Avancini, Menezes,
  Alloy, Marinelli, Moraes et~al.}}]{Avancini:2008zz}
\bibinfo{author}{\bibfnamefont{S.}~\bibnamefont{Avancini}},
  \bibinfo{author}{\bibfnamefont{D.}~\bibnamefont{Menezes}},
  \bibinfo{author}{\bibfnamefont{M.}~\bibnamefont{Alloy}},
  \bibinfo{author}{\bibfnamefont{J.}~\bibnamefont{Marinelli}},
  \bibinfo{author}{\bibfnamefont{M.}~\bibnamefont{Moraes}},
  \bibnamefont{et~al.}, \bibinfo{journal}{Phys. Rev.}
  \textbf{\bibinfo{volume}{C78}}, \bibinfo{pages}{015802}
  (\bibinfo{year}{2008}).

\bibitem[{\citenamefont{Avancini et~al.}(2009)\citenamefont{Avancini, Brito,
  Marinelli, Menezes, de~Moraes et~al.}}]{Avancini:2008kg}
\bibinfo{author}{\bibfnamefont{S.}~\bibnamefont{Avancini}},
  \bibinfo{author}{\bibfnamefont{L.}~\bibnamefont{Brito}},
  \bibinfo{author}{\bibfnamefont{J.}~\bibnamefont{Marinelli}},
  \bibinfo{author}{\bibfnamefont{D.}~\bibnamefont{Menezes}},
  \bibinfo{author}{\bibfnamefont{M.}~\bibnamefont{de~Moraes}},
  \bibnamefont{et~al.}, \bibinfo{journal}{Phys. Rev.}
  \textbf{\bibinfo{volume}{C79}}, \bibinfo{pages}{035804}
  (\bibinfo{year}{2009}).

\bibitem[{\citenamefont{Avancini et~al.}(2010)\citenamefont{Avancini,
  Chiacchiera, Menezes, and Providencia}}]{Avancini:2010ch}
\bibinfo{author}{\bibfnamefont{S.~S.} \bibnamefont{Avancini}},
  \bibinfo{author}{\bibfnamefont{S.}~\bibnamefont{Chiacchiera}},
  \bibinfo{author}{\bibfnamefont{D.~P.} \bibnamefont{Menezes}},
  \bibnamefont{and}
  \bibinfo{author}{\bibfnamefont{C.}~\bibnamefont{Providencia}},
  \bibinfo{journal}{Phys.Rev.} \textbf{\bibinfo{volume}{C82}},
  \bibinfo{pages}{055807} (\bibinfo{year}{2010}).

\bibitem[{\citenamefont{Avancini et~al.}(2012)\citenamefont{Avancini, Barros,
  Brito, Chiacchiera, Menezes, and Providencia}}]{Avancini:2012bj}
\bibinfo{author}{\bibfnamefont{S.~S.} \bibnamefont{Avancini}},
  \bibinfo{author}{\bibfnamefont{C.~C.} \bibnamefont{Barros},
  \bibfnamefont{Jr}}, \bibinfo{author}{\bibfnamefont{L.}~\bibnamefont{Brito}},
  \bibinfo{author}{\bibfnamefont{S.}~\bibnamefont{Chiacchiera}},
  \bibinfo{author}{\bibfnamefont{D.~P.} \bibnamefont{Menezes}},
  \bibnamefont{and}
  \bibinfo{author}{\bibfnamefont{C.}~\bibnamefont{Providencia}},
  \bibinfo{journal}{Phys. Rev.} \textbf{\bibinfo{volume}{C85}},
  \bibinfo{pages}{035806} (\bibinfo{year}{2012}).

\bibitem[{\citenamefont{Schneider et~al.}(2013)\citenamefont{Schneider,
  Horowitz, Hughto, and Berry}}]{Schneider:2013dwa}
\bibinfo{author}{\bibfnamefont{A.~S.} \bibnamefont{Schneider}},
  \bibinfo{author}{\bibfnamefont{C.~J.} \bibnamefont{Horowitz}},
  \bibinfo{author}{\bibfnamefont{J.}~\bibnamefont{Hughto}}, \bibnamefont{and}
  \bibinfo{author}{\bibfnamefont{D.~K.} \bibnamefont{Berry}},
  \bibinfo{journal}{Phys. Rev.} \textbf{\bibinfo{volume}{C88}},
  \bibinfo{pages}{065807} (\bibinfo{year}{2013}).

\bibitem[{\citenamefont{Caplan et~al.}(2015)\citenamefont{Caplan, Schneider,
  Horowitz, and Berry}}]{Caplan:2014gaa}
\bibinfo{author}{\bibfnamefont{M.~E.} \bibnamefont{Caplan}},
  \bibinfo{author}{\bibfnamefont{A.~S.} \bibnamefont{Schneider}},
  \bibinfo{author}{\bibfnamefont{C.~J.} \bibnamefont{Horowitz}},
  \bibnamefont{and} \bibinfo{author}{\bibfnamefont{D.~K.} \bibnamefont{Berry}},
  \bibinfo{journal}{Phys. Rev.} \textbf{\bibinfo{volume}{C91}},
  \bibinfo{pages}{065802} (\bibinfo{year}{2015}).

\bibitem[{\citenamefont{Horowitz et~al.}(2015)\citenamefont{Horowitz, Berry,
  Briggs, Caplan, Cumming, and Schneider}}]{Horowitz:2014xca}
\bibinfo{author}{\bibfnamefont{C.~J.} \bibnamefont{Horowitz}},
  \bibinfo{author}{\bibfnamefont{D.~K.} \bibnamefont{Berry}},
  \bibinfo{author}{\bibfnamefont{C.~M.} \bibnamefont{Briggs}},
  \bibinfo{author}{\bibfnamefont{M.~E.} \bibnamefont{Caplan}},
  \bibinfo{author}{\bibfnamefont{A.}~\bibnamefont{Cumming}}, \bibnamefont{and}
  \bibinfo{author}{\bibfnamefont{A.~S.} \bibnamefont{Schneider}},
  \bibinfo{journal}{Phys. Rev. Lett.} \textbf{\bibinfo{volume}{114}},
  \bibinfo{pages}{031102} (\bibinfo{year}{2015}).

\bibitem[{\citenamefont{Ravenhall et~al.}(1983)\citenamefont{Ravenhall,
  Pethick, and Wilson}}]{Ravenhall:1983uh}
\bibinfo{author}{\bibfnamefont{D.~G.} \bibnamefont{Ravenhall}},
  \bibinfo{author}{\bibfnamefont{C.~J.} \bibnamefont{Pethick}},
  \bibnamefont{and} \bibinfo{author}{\bibfnamefont{J.~R.}
  \bibnamefont{Wilson}}, \bibinfo{journal}{Phys. Rev. Lett.}
  \textbf{\bibinfo{volume}{50}}, \bibinfo{pages}{2066} (\bibinfo{year}{1983}).

\bibitem[{\citenamefont{Hashimoto et~al.}(1984)\citenamefont{Hashimoto, Seki,
  and Yamada}}]{Hashimoto:1984}
\bibinfo{author}{\bibfnamefont{M.}~\bibnamefont{Hashimoto}},
  \bibinfo{author}{\bibfnamefont{H.}~\bibnamefont{Seki}}, \bibnamefont{and}
  \bibinfo{author}{\bibfnamefont{M.}~\bibnamefont{Yamada}},
  \bibinfo{journal}{Prog. Theor. Phys.} \textbf{\bibinfo{volume}{71}},
  \bibinfo{pages}{320} (\bibinfo{year}{1984}).

\bibitem[{\citenamefont{Horowitz
  et~al.}(2004{\natexlab{a}})\citenamefont{Horowitz, Perez-Garcia, and
  Piekarewicz}}]{Horowitz:2004yf}
\bibinfo{author}{\bibfnamefont{C.~J.} \bibnamefont{Horowitz}},
  \bibinfo{author}{\bibfnamefont{M.~A.} \bibnamefont{Perez-Garcia}},
  \bibnamefont{and}
  \bibinfo{author}{\bibfnamefont{J.}~\bibnamefont{Piekarewicz}},
  \bibinfo{journal}{Phys. Rev.} \textbf{\bibinfo{volume}{C69}},
  \bibinfo{pages}{045804} (\bibinfo{year}{2004}{\natexlab{a}}).

\bibitem[{\citenamefont{Horowitz
  et~al.}(2004{\natexlab{b}})\citenamefont{Horowitz, Perez-Garcia, Carriere,
  Berry, and Piekarewicz}}]{Horowitz:2004pv}
\bibinfo{author}{\bibfnamefont{C.~J.} \bibnamefont{Horowitz}},
  \bibinfo{author}{\bibfnamefont{M.~A.} \bibnamefont{Perez-Garcia}},
  \bibinfo{author}{\bibfnamefont{J.}~\bibnamefont{Carriere}},
  \bibinfo{author}{\bibfnamefont{D.~K.} \bibnamefont{Berry}}, \bibnamefont{and}
  \bibinfo{author}{\bibfnamefont{J.}~\bibnamefont{Piekarewicz}},
  \bibinfo{journal}{Phys. Rev.} \textbf{\bibinfo{volume}{C70}},
  \bibinfo{pages}{065806} (\bibinfo{year}{2004}{\natexlab{b}}).

\bibitem[{\citenamefont{Horowitz et~al.}(2005)\citenamefont{Horowitz,
  Perez-Garcia, Berry, and Piekarewicz}}]{Horowitz:2005zb}
\bibinfo{author}{\bibfnamefont{C.~J.} \bibnamefont{Horowitz}},
  \bibinfo{author}{\bibfnamefont{M.~A.} \bibnamefont{Perez-Garcia}},
  \bibinfo{author}{\bibfnamefont{D.~K.} \bibnamefont{Berry}}, \bibnamefont{and}
  \bibinfo{author}{\bibfnamefont{J.}~\bibnamefont{Piekarewicz}},
  \bibinfo{journal}{Phys. Rev.} \textbf{\bibinfo{volume}{C72}},
  \bibinfo{pages}{035801} (\bibinfo{year}{2005}).

\bibitem[{\citenamefont{Watanabe et~al.}(2003)\citenamefont{Watanabe, Sato,
  Yasuoka, and Ebisuzaki}}]{Watanabe:2003xu}
\bibinfo{author}{\bibfnamefont{G.}~\bibnamefont{Watanabe}},
  \bibinfo{author}{\bibfnamefont{K.}~\bibnamefont{Sato}},
  \bibinfo{author}{\bibfnamefont{K.}~\bibnamefont{Yasuoka}}, \bibnamefont{and}
  \bibinfo{author}{\bibfnamefont{T.}~\bibnamefont{Ebisuzaki}},
  \bibinfo{journal}{Phys. Rev.} \textbf{\bibinfo{volume}{C68}},
  \bibinfo{pages}{035806} (\bibinfo{year}{2003}).

\bibitem[{\citenamefont{Watanabe et~al.}(2005)\citenamefont{Watanabe, Maruyama,
  Sato, Yasuoka, and Ebisuzaki}}]{Watanabe:2004tr}
\bibinfo{author}{\bibfnamefont{G.}~\bibnamefont{Watanabe}},
  \bibinfo{author}{\bibfnamefont{T.}~\bibnamefont{Maruyama}},
  \bibinfo{author}{\bibfnamefont{K.}~\bibnamefont{Sato}},
  \bibinfo{author}{\bibfnamefont{K.}~\bibnamefont{Yasuoka}}, \bibnamefont{and}
  \bibinfo{author}{\bibfnamefont{T.}~\bibnamefont{Ebisuzaki}},
  \bibinfo{journal}{Phys. Rev. Lett.} \textbf{\bibinfo{volume}{94}},
  \bibinfo{pages}{031101} (\bibinfo{year}{2005}).

\bibitem[{\citenamefont{Watanabe et~al.}(2009)\citenamefont{Watanabe, Sonoda,
  Maruyama, Sato, Yasuoka et~al.}}]{Watanabe:2009vi}
\bibinfo{author}{\bibfnamefont{G.}~\bibnamefont{Watanabe}},
  \bibinfo{author}{\bibfnamefont{H.}~\bibnamefont{Sonoda}},
  \bibinfo{author}{\bibfnamefont{T.}~\bibnamefont{Maruyama}},
  \bibinfo{author}{\bibfnamefont{K.}~\bibnamefont{Sato}},
  \bibinfo{author}{\bibfnamefont{K.}~\bibnamefont{Yasuoka}},
  \bibnamefont{et~al.}, \bibinfo{journal}{Phys. Rev. Lett.}
  \textbf{\bibinfo{volume}{103}}, \bibinfo{pages}{121101}
  (\bibinfo{year}{2009}).

\bibitem[{\citenamefont{Bulgac and Magierski}(2001)}]{Bulgac:2001}
\bibinfo{author}{\bibfnamefont{A.}~\bibnamefont{Bulgac}} \bibnamefont{and}
  \bibinfo{author}{\bibfnamefont{P.}~\bibnamefont{Magierski}},
  \bibinfo{journal}{Nuclear Physics A} \textbf{\bibinfo{volume}{683}},
  \bibinfo{pages}{695 } (\bibinfo{year}{2001}).

\bibitem[{\citenamefont{Magierski and Heenen}(2002)}]{Magierski:2001ud}
\bibinfo{author}{\bibfnamefont{P.}~\bibnamefont{Magierski}} \bibnamefont{and}
  \bibinfo{author}{\bibfnamefont{P.-H.} \bibnamefont{Heenen}},
  \bibinfo{journal}{Phys. Rev.} \textbf{\bibinfo{volume}{C65}},
  \bibinfo{pages}{045804} (\bibinfo{year}{2002}).

\bibitem[{\citenamefont{Chamel}(2005)}]{Chamel:2004in}
\bibinfo{author}{\bibfnamefont{N.}~\bibnamefont{Chamel}},
  \bibinfo{journal}{Nucl. Phys.} \textbf{\bibinfo{volume}{A747}},
  \bibinfo{pages}{109} (\bibinfo{year}{2005}).

\bibitem[{\citenamefont{Newton and Stone}(2009)}]{Newton:2009zz}
\bibinfo{author}{\bibfnamefont{W.}~\bibnamefont{Newton}} \bibnamefont{and}
  \bibinfo{author}{\bibfnamefont{J.}~\bibnamefont{Stone}},
  \bibinfo{journal}{Phys. Rev.} \textbf{\bibinfo{volume}{C79}},
  \bibinfo{pages}{055801} (\bibinfo{year}{2009}).

\bibitem[{\citenamefont{Schuetrumpf and
  Nazarewicz}(2015)}]{Schuetrumpf:2015nza}
\bibinfo{author}{\bibfnamefont{B.}~\bibnamefont{Schuetrumpf}} \bibnamefont{and}
  \bibinfo{author}{\bibfnamefont{W.}~\bibnamefont{Nazarewicz}},
  \bibinfo{journal}{Phys. Rev.} \textbf{\bibinfo{volume}{C92}},
  \bibinfo{pages}{045806} (\bibinfo{year}{2015}).

\bibitem[{\citenamefont{Chamel and Haensel}(2008)}]{Chamel:2008ca}
\bibinfo{author}{\bibfnamefont{N.}~\bibnamefont{Chamel}} \bibnamefont{and}
  \bibinfo{author}{\bibfnamefont{P.}~\bibnamefont{Haensel}},
  \bibinfo{journal}{Living Rev. Rel.} \textbf{\bibinfo{volume}{11}},
  \bibinfo{pages}{10} (\bibinfo{year}{2008}), \eprint{0812.3955}.

\bibitem[{\citenamefont{Bertulani and Piekarewicz}(2012)}]{Bertulani:2012}
\bibinfo{author}{\bibfnamefont{C.}~\bibnamefont{Bertulani}} \bibnamefont{and}
  \bibinfo{author}{\bibfnamefont{J.}~\bibnamefont{Piekarewicz}},
  \emph{\bibinfo{title}{Neutron Star Crust.}} (\bibinfo{publisher}{Nova Science
  Publishers, Hauppauge New York}, \bibinfo{year}{2012}).

\bibitem[{\citenamefont{Todd-Rutel and Piekarewicz}(2005)}]{Todd-Rutel:2005fa}
\bibinfo{author}{\bibfnamefont{B.~G.} \bibnamefont{Todd-Rutel}}
  \bibnamefont{and}
  \bibinfo{author}{\bibfnamefont{J.}~\bibnamefont{Piekarewicz}},
  \bibinfo{journal}{Phys. Rev. Lett} \textbf{\bibinfo{volume}{95}},
  \bibinfo{pages}{122501} (\bibinfo{year}{2005}).

\bibitem[{\citenamefont{Fattoyev and Piekarewicz}(2013)}]{Fattoyev:2013yaa}
\bibinfo{author}{\bibfnamefont{F.}~\bibnamefont{Fattoyev}} \bibnamefont{and}
  \bibinfo{author}{\bibfnamefont{J.}~\bibnamefont{Piekarewicz}},
  \bibinfo{journal}{Phys. Rev. Lett.} \textbf{\bibinfo{volume}{111}},
  \bibinfo{pages}{162501} (\bibinfo{year}{2013}).

\bibitem[{\citenamefont{Ozel et~al.}(2010)\citenamefont{Ozel, Baym, and
  Guver}}]{Ozel:2010fw}
\bibinfo{author}{\bibfnamefont{F.}~\bibnamefont{Ozel}},
  \bibinfo{author}{\bibfnamefont{G.}~\bibnamefont{Baym}}, \bibnamefont{and}
  \bibinfo{author}{\bibfnamefont{T.}~\bibnamefont{Guver}},
  \bibinfo{journal}{Phys. Rev.} \textbf{\bibinfo{volume}{D82}},
  \bibinfo{pages}{101301} (\bibinfo{year}{2010}).

\bibitem[{\citenamefont{Steiner et~al.}(2010)\citenamefont{Steiner, Lattimer,
  and Brown}}]{Steiner:2010fz}
\bibinfo{author}{\bibfnamefont{A.~W.} \bibnamefont{Steiner}},
  \bibinfo{author}{\bibfnamefont{J.~M.} \bibnamefont{Lattimer}},
  \bibnamefont{and} \bibinfo{author}{\bibfnamefont{E.~F.} \bibnamefont{Brown}},
  \bibinfo{journal}{Astrophys. J.} \textbf{\bibinfo{volume}{722}},
  \bibinfo{pages}{33} (\bibinfo{year}{2010}).

\bibitem[{\citenamefont{Suleimanov et~al.}(2011)\citenamefont{Suleimanov,
  Poutanen, Revnivtsev, and Werner}}]{Suleimanov:2010th}
\bibinfo{author}{\bibfnamefont{V.}~\bibnamefont{Suleimanov}},
  \bibinfo{author}{\bibfnamefont{J.}~\bibnamefont{Poutanen}},
  \bibinfo{author}{\bibfnamefont{M.}~\bibnamefont{Revnivtsev}},
  \bibnamefont{and} \bibinfo{author}{\bibfnamefont{K.}~\bibnamefont{Werner}},
  \bibinfo{journal}{Astrophys. J.} \textbf{\bibinfo{volume}{742}},
  \bibinfo{pages}{122} (\bibinfo{year}{2011}).

\bibitem[{\citenamefont{Guillot et~al.}(2013)\citenamefont{Guillot, Servillat,
  Webb, and Ruledge}}]{Guillot:2013wu}
\bibinfo{author}{\bibfnamefont{S.}~\bibnamefont{Guillot}},
  \bibinfo{author}{\bibfnamefont{M.}~\bibnamefont{Servillat}},
  \bibinfo{author}{\bibfnamefont{N.~A.} \bibnamefont{Webb}}, \bibnamefont{and}
  \bibinfo{author}{\bibfnamefont{R.~E.} \bibnamefont{Ruledge}},
  \bibinfo{journal}{Astrophys.J.} \textbf{\bibinfo{volume}{772}},
  \bibinfo{pages}{7} (\bibinfo{year}{2013}).

\bibitem[{\citenamefont{Lattimer and Steiner}(2014)}]{Lattimer:2013hma}
\bibinfo{author}{\bibfnamefont{J.~M.} \bibnamefont{Lattimer}} \bibnamefont{and}
  \bibinfo{author}{\bibfnamefont{A.~W.} \bibnamefont{Steiner}},
  \bibinfo{journal}{Astrophys. J.} \textbf{\bibinfo{volume}{784}},
  \bibinfo{pages}{123} (\bibinfo{year}{2014}).

\bibitem[{\citenamefont{Heinke et~al.}(2014)\citenamefont{Heinke, Cohn, Lugger,
  Webb, Ho et~al.}}]{Heinke:2014xaa}
\bibinfo{author}{\bibfnamefont{C.~O.} \bibnamefont{Heinke}},
  \bibinfo{author}{\bibfnamefont{H.~N.} \bibnamefont{Cohn}},
  \bibinfo{author}{\bibfnamefont{P.~M.} \bibnamefont{Lugger}},
  \bibinfo{author}{\bibfnamefont{N.~A.} \bibnamefont{Webb}},
  \bibinfo{author}{\bibfnamefont{W.}~\bibnamefont{Ho}}, \bibnamefont{et~al.},
  \bibinfo{journal}{Mon. Not. Roy. Astron. Soc.}
  \textbf{\bibinfo{volume}{444}}, \bibinfo{pages}{443} (\bibinfo{year}{2014}).

\bibitem[{\citenamefont{Guillot and Rutledge}(2014)}]{Guillot:2014lla}
\bibinfo{author}{\bibfnamefont{S.}~\bibnamefont{Guillot}} \bibnamefont{and}
  \bibinfo{author}{\bibfnamefont{R.~E.} \bibnamefont{Rutledge}},
  \bibinfo{journal}{Astrophys.J.} \textbf{\bibinfo{volume}{796}},
  \bibinfo{pages}{L3} (\bibinfo{year}{2014}).

\bibitem[{\citenamefont{Ozel et~al.}(2016)\citenamefont{Ozel, Psaltis, Guver,
  Baym, Heinke, and Guillot}}]{Ozel:2015fia}
\bibinfo{author}{\bibfnamefont{F.}~\bibnamefont{Ozel}},
  \bibinfo{author}{\bibfnamefont{D.}~\bibnamefont{Psaltis}},
  \bibinfo{author}{\bibfnamefont{T.}~\bibnamefont{Guver}},
  \bibinfo{author}{\bibfnamefont{G.}~\bibnamefont{Baym}},
  \bibinfo{author}{\bibfnamefont{C.}~\bibnamefont{Heinke}}, \bibnamefont{and}
  \bibinfo{author}{\bibfnamefont{S.}~\bibnamefont{Guillot}},
  \bibinfo{journal}{Astrophys. J.} \textbf{\bibinfo{volume}{820}},
  \bibinfo{pages}{28} (\bibinfo{year}{2016}).

\bibitem[{\citenamefont{Watts et~al.}(2016)}]{Watts:2016uzu}
\bibinfo{author}{\bibfnamefont{A.~L.} \bibnamefont{Watts}}
  \bibnamefont{et~al.}, \bibinfo{journal}{Rev. Mod. Phys.}
  \textbf{\bibinfo{volume}{88}}, \bibinfo{pages}{021001}
  (\bibinfo{year}{2016}).

\bibitem[{\citenamefont{Steiner et~al.}(2018)\citenamefont{Steiner, Heinke,
  Bogdanov, Li, Ho, Bahramian, and Han}}]{Steiner:2017vmg}
\bibinfo{author}{\bibfnamefont{A.~W.} \bibnamefont{Steiner}},
  \bibinfo{author}{\bibfnamefont{C.~O.} \bibnamefont{Heinke}},
  \bibinfo{author}{\bibfnamefont{S.}~\bibnamefont{Bogdanov}},
  \bibinfo{author}{\bibfnamefont{C.}~\bibnamefont{Li}},
  \bibinfo{author}{\bibfnamefont{W.~C.~G.} \bibnamefont{Ho}},
  \bibinfo{author}{\bibfnamefont{A.}~\bibnamefont{Bahramian}},
  \bibnamefont{and} \bibinfo{author}{\bibfnamefont{S.}~\bibnamefont{Han}},
  \bibinfo{journal}{Mon. Not. Roy. Astron. Soc.}
  \textbf{\bibinfo{volume}{476}}, \bibinfo{pages}{421} (\bibinfo{year}{2018}).

\bibitem[{\citenamefont{N\"attil\"a et~al.}(2017)\citenamefont{N\"attil\"a,
  Miller, Steiner, Kajava, Suleimanov, and Poutanen}}]{Nattila:2017wtj}
\bibinfo{author}{\bibfnamefont{J.}~\bibnamefont{N\"attil\"a}},
  \bibinfo{author}{\bibfnamefont{M.~C.} \bibnamefont{Miller}},
  \bibinfo{author}{\bibfnamefont{A.~W.} \bibnamefont{Steiner}},
  \bibinfo{author}{\bibfnamefont{J.~J.~E.} \bibnamefont{Kajava}},
  \bibinfo{author}{\bibfnamefont{V.~F.} \bibnamefont{Suleimanov}},
  \bibnamefont{and} \bibinfo{author}{\bibfnamefont{J.}~\bibnamefont{Poutanen}},
  \bibinfo{journal}{Astron. Astrophys.} \textbf{\bibinfo{volume}{608}},
  \bibinfo{pages}{A31} (\bibinfo{year}{2017}).

\bibitem[{\citenamefont{Lattimer and Prakash}(2007)}]{Lattimer:2006xb}
\bibinfo{author}{\bibfnamefont{J.~M.} \bibnamefont{Lattimer}} \bibnamefont{and}
  \bibinfo{author}{\bibfnamefont{M.}~\bibnamefont{Prakash}},
  \bibinfo{journal}{Phys. Rept.} \textbf{\bibinfo{volume}{442}},
  \bibinfo{pages}{109} (\bibinfo{year}{2007}).

\bibitem[{\citenamefont{Piekarewicz and Centelles}(2009)}]{Piekarewicz:2008nh}
\bibinfo{author}{\bibfnamefont{J.}~\bibnamefont{Piekarewicz}} \bibnamefont{and}
  \bibinfo{author}{\bibfnamefont{M.}~\bibnamefont{Centelles}},
  \bibinfo{journal}{Phys. Rev.} \textbf{\bibinfo{volume}{C79}},
  \bibinfo{pages}{054311} (\bibinfo{year}{2009}).

\bibitem[{\citenamefont{Brown}(2000)}]{Brown:2000}
\bibinfo{author}{\bibfnamefont{B.~A.} \bibnamefont{Brown}},
  \bibinfo{journal}{Phys. Rev. Lett.} \textbf{\bibinfo{volume}{85}},
  \bibinfo{pages}{5296} (\bibinfo{year}{2000}).

\bibitem[{\citenamefont{Furnstahl}(2002)}]{Furnstahl:2001un}
\bibinfo{author}{\bibfnamefont{R.~J.} \bibnamefont{Furnstahl}},
  \bibinfo{journal}{Nucl. Phys.} \textbf{\bibinfo{volume}{A706}},
  \bibinfo{pages}{85} (\bibinfo{year}{2002}).

\bibitem[{\citenamefont{Centelles et~al.}(2009)\citenamefont{Centelles,
  Roca-Maza, Vi\~nas, and Warda}}]{Centelles:2008vu}
\bibinfo{author}{\bibfnamefont{M.}~\bibnamefont{Centelles}},
  \bibinfo{author}{\bibfnamefont{X.}~\bibnamefont{Roca-Maza}},
  \bibinfo{author}{\bibfnamefont{X.}~\bibnamefont{Vi\~nas}}, \bibnamefont{and}
  \bibinfo{author}{\bibfnamefont{M.}~\bibnamefont{Warda}},
  \bibinfo{journal}{Phys. Rev. Lett.} \textbf{\bibinfo{volume}{102}},
  \bibinfo{pages}{122502} (\bibinfo{year}{2009}).

\bibitem[{\citenamefont{Roca-Maza
  et~al.}(2011{\natexlab{b}})\citenamefont{Roca-Maza, Centelles, Vi\~nas, and
  Warda}}]{RocaMaza:2011pm}
\bibinfo{author}{\bibfnamefont{X.}~\bibnamefont{Roca-Maza}},
  \bibinfo{author}{\bibfnamefont{M.}~\bibnamefont{Centelles}},
  \bibinfo{author}{\bibfnamefont{X.}~\bibnamefont{Vi\~nas}}, \bibnamefont{and}
  \bibinfo{author}{\bibfnamefont{M.}~\bibnamefont{Warda}},
  \bibinfo{journal}{Phys. Rev. Lett.} \textbf{\bibinfo{volume}{106}},
  \bibinfo{pages}{252501} (\bibinfo{year}{2011}{\natexlab{b}}).

\bibitem[{\citenamefont{Horowitz and
  Piekarewicz}(2001{\natexlab{a}})}]{Horowitz:2000xj}
\bibinfo{author}{\bibfnamefont{C.~J.} \bibnamefont{Horowitz}} \bibnamefont{and}
  \bibinfo{author}{\bibfnamefont{J.}~\bibnamefont{Piekarewicz}},
  \bibinfo{journal}{Phys. Rev. Lett.} \textbf{\bibinfo{volume}{86}},
  \bibinfo{pages}{5647} (\bibinfo{year}{2001}{\natexlab{a}}).

\bibitem[{\citenamefont{Horowitz and
  Piekarewicz}(2001{\natexlab{b}})}]{Horowitz:2001ya}
\bibinfo{author}{\bibfnamefont{C.~J.} \bibnamefont{Horowitz}} \bibnamefont{and}
  \bibinfo{author}{\bibfnamefont{J.}~\bibnamefont{Piekarewicz}},
  \bibinfo{journal}{Phys. Rev.} \textbf{\bibinfo{volume}{C64}},
  \bibinfo{pages}{062802} (\bibinfo{year}{2001}{\natexlab{b}}).

\bibitem[{\citenamefont{Carriere et~al.}(2003)\citenamefont{Carriere, Horowitz,
  and Piekarewicz}}]{Carriere:2002bx}
\bibinfo{author}{\bibfnamefont{J.}~\bibnamefont{Carriere}},
  \bibinfo{author}{\bibfnamefont{C.~J.} \bibnamefont{Horowitz}},
  \bibnamefont{and}
  \bibinfo{author}{\bibfnamefont{J.}~\bibnamefont{Piekarewicz}},
  \bibinfo{journal}{Astrophys. J.} \textbf{\bibinfo{volume}{593}},
  \bibinfo{pages}{463} (\bibinfo{year}{2003}).

\bibitem[{\citenamefont{Fattoyev and Piekarewicz}(2012)}]{Fattoyev:2012rm}
\bibinfo{author}{\bibfnamefont{F.}~\bibnamefont{Fattoyev}} \bibnamefont{and}
  \bibinfo{author}{\bibfnamefont{J.}~\bibnamefont{Piekarewicz}},
  \bibinfo{journal}{Phys. Rev.} \textbf{\bibinfo{volume}{C88}}
  (\bibinfo{year}{2012}).

\bibitem[{\citenamefont{Chen and Piekarewicz}(2014)}]{Chen:2014sca}
\bibinfo{author}{\bibfnamefont{W.-C.} \bibnamefont{Chen}} \bibnamefont{and}
  \bibinfo{author}{\bibfnamefont{J.}~\bibnamefont{Piekarewicz}},
  \bibinfo{journal}{Phys. Rev.} \textbf{\bibinfo{volume}{C90}},
  \bibinfo{pages}{044305} (\bibinfo{year}{2014}).

\bibitem[{\citenamefont{Abrahamyan et~al.}(2012)\citenamefont{Abrahamyan,
  Ahmed, Albataineh, Aniol, Armstrong et~al.}}]{Abrahamyan:2012gp}
\bibinfo{author}{\bibfnamefont{S.}~\bibnamefont{Abrahamyan}},
  \bibinfo{author}{\bibfnamefont{Z.}~\bibnamefont{Ahmed}},
  \bibinfo{author}{\bibfnamefont{H.}~\bibnamefont{Albataineh}},
  \bibinfo{author}{\bibfnamefont{K.}~\bibnamefont{Aniol}},
  \bibinfo{author}{\bibfnamefont{D.~S.} \bibnamefont{Armstrong}},
  \bibnamefont{et~al.}, \bibinfo{journal}{Phys. Rev. Lett.}
  \textbf{\bibinfo{volume}{108}}, \bibinfo{pages}{112502}
  (\bibinfo{year}{2012}).

\bibitem[{\citenamefont{Horowitz et~al.}(2012)\citenamefont{Horowitz, Ahmed,
  Jen, Rakhman, Souder et~al.}}]{Horowitz:2012tj}
\bibinfo{author}{\bibfnamefont{C.~J.} \bibnamefont{Horowitz}},
  \bibinfo{author}{\bibfnamefont{Z.}~\bibnamefont{Ahmed}},
  \bibinfo{author}{\bibfnamefont{C.~M.} \bibnamefont{Jen}},
  \bibinfo{author}{\bibfnamefont{A.}~\bibnamefont{Rakhman}},
  \bibinfo{author}{\bibfnamefont{P.~A.} \bibnamefont{Souder}},
  \bibnamefont{et~al.}, \bibinfo{journal}{Phys. Rev.}
  \textbf{\bibinfo{volume}{C85}}, \bibinfo{pages}{032501}
  (\bibinfo{year}{2012}).

\bibitem[{\citenamefont{Damour et~al.}(1992)\citenamefont{Damour, Soffel, and
  Xu}}]{Damour:1991yw}
\bibinfo{author}{\bibfnamefont{T.}~\bibnamefont{Damour}},
  \bibinfo{author}{\bibfnamefont{M.}~\bibnamefont{Soffel}}, \bibnamefont{and}
  \bibinfo{author}{\bibfnamefont{C.-m.} \bibnamefont{Xu}},
  \bibinfo{journal}{Phys. Rev.} \textbf{\bibinfo{volume}{D45}},
  \bibinfo{pages}{1017} (\bibinfo{year}{1992}).

\bibitem[{\citenamefont{Flanagan and Hinderer}(2008)}]{Flanagan:2007ix}
\bibinfo{author}{\bibfnamefont{E.~E.} \bibnamefont{Flanagan}} \bibnamefont{and}
  \bibinfo{author}{\bibfnamefont{T.}~\bibnamefont{Hinderer}},
  \bibinfo{journal}{Phys. Rev.} \textbf{\bibinfo{volume}{D77}},
  \bibinfo{pages}{021502} (\bibinfo{year}{2008}).

\bibitem[{\citenamefont{Fattoyev et~al.}(2018)\citenamefont{Fattoyev,
  Piekarewicz, and Horowitz}}]{Fattoyev:2017jql}
\bibinfo{author}{\bibfnamefont{F.~J.} \bibnamefont{Fattoyev}},
  \bibinfo{author}{\bibfnamefont{J.}~\bibnamefont{Piekarewicz}},
  \bibnamefont{and} \bibinfo{author}{\bibfnamefont{C.~J.}
  \bibnamefont{Horowitz}}, \bibinfo{journal}{Phys. Rev. Lett.}
  \textbf{\bibinfo{volume}{120}}, \bibinfo{pages}{172702}
  (\bibinfo{year}{2018}).

\bibitem[{\citenamefont{Fattoyev et~al.}(2010)\citenamefont{Fattoyev, Horowitz,
  Piekarewicz, and Shen}}]{Fattoyev:2010mx}
\bibinfo{author}{\bibfnamefont{F.~J.} \bibnamefont{Fattoyev}},
  \bibinfo{author}{\bibfnamefont{C.~J.} \bibnamefont{Horowitz}},
  \bibinfo{author}{\bibfnamefont{J.}~\bibnamefont{Piekarewicz}},
  \bibnamefont{and} \bibinfo{author}{\bibfnamefont{G.}~\bibnamefont{Shen}},
  \bibinfo{journal}{Phys. Rev.} \textbf{\bibinfo{volume}{C82}},
  \bibinfo{pages}{055803} (\bibinfo{year}{2010}).

\bibitem[{nat(2012)}]{national2012Nuclear}
\emph{\bibinfo{title}{Nuclear Physics: Exploring the Heart of Matter}}
  (\bibinfo{publisher}{The National Academies Press},
  \bibinfo{address}{Washington}, \bibinfo{year}{2012}).

\end{thebibliography}

\vfill\eject
\end{document}